\documentclass[11pt]{article}

\usepackage{amsmath,amsthm,amsfonts,amssymb,amsopn,amscd, mathrsfs}
\usepackage{graphics}
\usepackage{latexsym}

\newtheorem{thm}{Theorem}[section]
\newtheorem{lem}[thm]{Lemma}
\newtheorem{cor}[thm]{Corollary}
\newtheorem{prp}[thm]{Proposition}

\theoremstyle{definition}

\newtheorem{exm}{Example}
\theoremstyle{remark}
\newtheorem{rmk}[thm]{Remark} 

\def\K{\mathcal{K}}
\def\vect{\mathrm{Vect(S^1)}}

\def\witt{\mathrm{Witt}}
\def\vir{\mathrm{Vir}}
\def\C{\mathbb{C}}
\def\R{\mathbb{R}}
\def\Z{\mathbb{Z}}

\title{Representation theory of the stabilizer subgroup of the point at infinity in 
$\mathrm{Diff}(S^1)$}
\date{}

\author{
{\bf Yoh Tanimoto\footnote{Supported in part by the ERC Advanced Grant 227458
OACFT ``Operator Algebras and Conformal Field Theory''.}}\\
Dipartimento di Matematica, Universit\`a di Roma ``Tor
Vergata''\\ Via della Ricerca Scientifica, 1 - I--00133 Roma, Italy.\\
E-mail: {\tt tanimoto@mat.uniroma2.it}}

\begin{document}
\maketitle
\begin{abstract}
The group $\mathrm{Diff}(S^1)$ of the orientation preserving diffeomorphisms
of the circle $S^1$ plays an important role in conformal field theory. We consider a subgroup
$B_0$ of $\mathrm{Diff}(S^1)$ whose elements stabilize ``the point at infinity''.
This subgroup is of interest for the actual physical theory living on the punctured circle,
or the real line.

We investigate the unique central extension $\mathcal{K}$ of the Lie algebra of that group.
We determine the first and second cohomologies, its ideal structure and the automorphism
group. We define a generalization of Verma modules and determine when these
representations are irreducible. Its endomorphism semigroup is investigated
and some unitary representations of the group which do not extend to $\mathrm{Diff}(S^1)$
are constructed.
\end{abstract}

\section{Introduction}
In this paper we study a certain subalgebra of the Virasoro algebra defined below. The Virasoro algebra
is a fundamental object in conformal quantum field theory.

The symmetry group of the chiral component of a conformal field theory in 1+1 dimension is $B_0$,
the group of all orientation-preserving diffeomorphisms of the real line
which are smooth at the point at infinity
(for example, see \cite{schottenloher}). Instead of working on $\mathbb{R}$, 
it is customary to consider a chiral model on the compactified line $S^1$
with the symmetry group
$\mathrm{Diff}(S^1)$. In a quantum theory, we are interested in its projective representations.

With positivity of the energy, which is a physical requirement, the representation
theory of the central extension of $\mathrm{Diff}(S^1)$ has been well studied
\cite{schottenloher}.
In any irreducible unitary projective representation of $\mathrm{Diff}(S^1)$, the central
element acts as a scalar $c$. The (central extension of the) group $\mathrm{Diff}(S^1)$
has a subgroup $S^1$ of rotations
and by positivity of energy the subgroup has the lowest eigenvalue $h \ge 0$. It is known
for which values of $c$ and $h$ there exist irreducible, unitary,
positive-energy, projective representations of $\mathrm{Diff}(S^1)$.
All such representations are classified by $c$ and $h$.

The Lie algebra of $\mathrm{Diff}(S^1)$ is the algebra of all the smooth vector fields
on $S^1$ \cite{milnor}.
It is sometimes convenient to study its polynomial subalgebra, the Witt algebra.
The Witt algebra has a unique central extension \cite{schottenloher}
called the Virasoro algebra
$\mathrm{Vir}$.
In a similar way as above, we can define lowest energy representations of $\mathrm{Vir}$ with
parameters $c,h$ and it is known when these representations are
unitary \cite{KR}. On the other hand, for any positive energy, unitary lowest weight 
representation
of $\mathrm{Vir}$ there is a corresponding projective representation of $\mathrm{Diff}(S^1)$ \cite{GW}.

In a physical context, conformal field theory in 1+1 dimensional Minkowski space can be
decomposed into its chiral components on two lightlines.
Thus it is mathematically useful to study the subgroup $B_0$
of stabilizers of one point (``the point at infinity'') of
$\mathrm{Diff}(S^1)$. We can construct nets of von Neumann algebras on $\mathbb{R}$
from representations of $B_0$, and nets on $\mathbb{R}^2$ by tensor product.
The theory of local quantum physics are extensively studied with techniques of von Neumann
algebras \cite{haag}\cite{borchers}\cite{lechner}\cite{kawahigashi}.
In the case of nets on $S^1$, the nets generated
by $\mathrm{Diff}(S^1)$ play a key role in the classification of diffeomorphism
covariant nets \cite{kl}.
This gives a strong motivation for studying the 
representation theory of $B_0$, since for nets on $\mathbb{R}$ the group $B_0$ should 
play a similar role to that of $\mathrm{Diff}(S^1)$ for nets on $S^1$.

Some properties of the restrictions of representations of
$\mathrm{Diff}(S^1)$ to $B_0$ have been studied. For example, the restriction to $B_0$
of every irreducible unitary positive energy representation of $\mathrm{Diff}(S^1)$ is
irreducible \cite{weiner}.
Different values of $c,h$ may correspond to equivalent representations \cite{weiner}.
Unfortunately little is known about representations which are not
restrictions. In this paper we address this problem.

\subsection{Preliminaries}
We identify the real line with the punctured circle through the Cayley transformation:
\[
x = i\frac{1+z}{1-z} \Longleftrightarrow z = \frac{x-i}{x+i}, x \in \R, z \in S^1 \subset \C.
\]
The group $\mathrm{Diff}(S^1)$ contains the following important one-parameter subgroups.
They are called respectively the groups of rotations, translations and dilations:
\begin{eqnarray*}
\rho_s(z) &=& e^{is}z, \mbox{ for } z \in S^1 \subset \mathbb{C}\\
\tau_s(x) &=& x + s, \mbox{ for } x \in \mathbb{R} \\
\delta_s(x) &=& e^{s}x, \mbox{ for } x \in \mathbb{R},
\end{eqnarray*}
where rotations are defined in the circle picture, on the other hand translations and 
dilations are defined in the real line picture.
Here we see that the point $z = e^{2\pi i\theta} = 1$ or $\theta = 0$ on the circle is identified
with the point at infinity in the real line picture.

The positivity of the energy for $\mathrm{Diff}(S^1)$ is usually defined as the boundedness
from below of the generator of the group of rotation (since we consider projective representations,
the generator of a one-parameter subgroup is defined only up to an addition of a real scalar
multiple of the identity).
It is well known that this is equivalent to the
boundedness from below of the generator of the group of translations (see \cite{longo}). The latter 
definition is the one having its origin in physics. Concerning the group $B_0$, as it
does not include the group of rotations, the positivity of energy is defined by
boundedness from below of the generator of the group of translations.

In the rest of this section we explain our notation regarding some infinite dimensional Lie
algebras (see \cite{schottenloher}).

The Witt algebra (we denote it by $\mathrm{Witt}$) is the Lie algebra generated by
$L_n$ for $n \in \mathbb{Z}$ with the following commutation relations:
\[
  [L_m, L_n] = (m-n)L_{m+n}.
\]
The Witt algebra has a central extension with a central element $C$, unique up to isomorphisms,
with the following commutation relations:
\[
  [L_m, L_n] = (m-n)L_{m+n} + \frac{C}{12}m(m^2-1)\delta_{m,-n}.
\]
This algebra is called the Virasoro algebra $\mathrm{Vir}$. On $\mathrm{Witt}$ and $\mathrm{Vir}$ we can define
a *-operation by
\[
(L_n)^* = L_{-n}, C^* = C.
\]

The Witt algebra is a subalgebra of the Lie algebra $\mathrm{Vect}(S^1)$ of smooth complex functions on the circle $S^1$
with the following commutation relations:
\[
  [f,g] = f g^\prime - f^\prime g,
\]
and the correspondence $L_n \mapsto ie^{in\theta}$. Its real part is the Lie algebra of the
group of diffeomorphisms of $S^1$\cite{milnor}. This algebra is equipped with the smooth
topology, namely, a net of functions $f_n$ converges to $f$ if and only if
the $k$-th derivatives $f_n^{(k)}$ converge to $f^{(k)}$ uniformly on $S^1$ for all
$k \ge 0$. The central extension above extends continuously to this algebra.
As the group $\mathrm{Diff}(S^1)$ is a manifold modelled on $\mathrm{Vect}(S^1)$,
it is equipped with the induced topology of the smooth topology of $\mathrm{Vect}(S^1)$.

We consider a subspace $\mathcal{K}_0$ of the Witt algebra spanned by $K_n = L_n - L_0$
for $n \neq 0$. By a straightforward calculation this subspace is indeed a *-subalgebra
with the following commutation relations:
\[
  [K_m, K_n] = \left\{
  \begin{array}{cc}
    (m-n)K_{m+n} - mK_m + nK_n & (m \neq -n) \\
    - mK_m - mK_{-m} & (m = -n)
  \end{array}
\right. .
\]

We denote $\vect_0 \subset \vect$ the subalgebra of smooth functions which vanish on
$\theta = 0$. This is the Lie algebra of the group $B_0$ of all
the diffeomorphisms of $S^1$ which stabilize $\theta = 0$.
The algebra $\K_0$ is a *-subalgebra of $\mathrm{Vect}(S^1)_0$.

We will show that $\mathcal{K}_0$ has a unique (up to isomorphisms) central extension
which is a subalgebra of Vir.
The central extension is denoted by $\mathcal{K}$ and has the following commutation
relations:
\begin{eqnarray}\label{kcomm}
  [K_m, K_n] &=& \left\{
  \begin{array}{cc}
    (m-n)K_{m+n} - mK_m + nK_n & (m \neq -n) \\
    - mK_m - mK_{-m} + \frac{C}{12}m(m^2-1) & (m = -n)
  \end{array}
\right. .
\end{eqnarray}

In section \ref{cohomology}, we determine the first and second cohomologies
of the algebra $\mathcal{K}_0$. The first cohomology corresponds to
one dimensional representations and the second cohomology corresponds
to central extensions. It will be shown that the only possible central
extension is the natural inclusion into the Virasoro algebra. On the other hand
the first cohomology is one dimensional and does not extend to
$\mathrm{Vir}$.

In section \ref{derivedsubalgebras}, we determine the ideal structure
of $\mathcal{K}_0$ and calculate their commutator subalgebras. It will
be shown that all of these ideals can be defined by the vanishing of certain
derivatives at the point at infinity.

In section \ref{theautomorphism}, we determine the automorphism group
of the central extension $\mathcal{K}$ of $\mathcal{K}_0$. This group
turns out to be very small but contains some elements not
extending to automorphisms of the Virasoro algebra.

In section \ref{generalizedverma}, we construct several
representations of $\mathcal{K}$. Each of these representations
has an analogue of a lowest weight vector and has the universal
property. Thanks to the result of Feigin and Fuks \cite{FF},
we can completely determine which of these representations are irreducible.

In section \ref{endomorphismsof}, we investigate the endomorphism semigroup
of $\mathcal{K}$. Compositions of these endomorphisms with known
unitary representations give rise some strange kinds of representations.
Corresponding representations of the group $B_0$ are
studied in section \ref{someunitary}.

\section{First and Second cohomologies of $\mathcal{K}_0$}\label{cohomology}
We will discuss the following cohomology groups of $\mathcal{K}_0$ \cite{schottenloher}:
\begin{eqnarray*}
  H^1(\mathcal{K}_0, \mathbb{C}) &:=& \{\phi:\mathcal{K}_0 \to \mathbb{C}| \mbox{ $\phi$ is
                              linear and vanishes on $[\mathcal{K}_0, \mathcal{K}_0]$.}\} \\
  Z^2(\mathcal{K}_0, \mathbb{C}) &:=& \{\omega:\mathcal{K}_0 \times \mathcal{K}_0
   \to \mathbb{C}| \mbox{ $\omega$ is bilinear and} \\
 & & \mbox{ for } a,b,c \in \mathcal{K}_0 \mbox{ satisfies }\omega(a,b) = -\omega(b,a), \\
 & & \mbox{ } \omega([a,b],c) + \omega([b,c],a) + \omega([c,a],b) = 0 \}\\
  B^2(\mathcal{K}_0, \mathbb{C}) &:=& \{\omega:\mathcal{K}_0 \times \mathcal{K}_0
   \to \mathbb{C}| \mbox{ there is $\mu$ s.t $\omega(a,b) = \mu([a,b])$}.\} \\
  H^2(\mathcal{K}_0, \mathbb{C}) &:=& Z^2/B^2.
\end{eqnarray*}

Elements in the (additive) group $H^1$ correspond to one dimensional representations of
$\mathcal{K}_0$. The group $H^2$ corresponds to the set of all central extensions
of $\mathcal{K}_0$. We call $H^1$ and $H^2$ the first and the second cohomology groups
of $\mathcal{K}_0$, respectively.

\begin{lem}\label{first}
$[\mathcal{K}_0, \mathcal{K}_0]$ has codimension one in $\mathcal{K}_0$.
\end{lem}
\begin{proof}
Let us define a linear functional $\phi$ on $\mathcal{K}_0$ by the following:
\[
  \phi(K_n) = n.
\]
As $K_n$'s form a basis of $\mathcal{K}_0$, this defines a linear functional.
By the commutation relation above, we have
\[
  \phi([K_m, K_n]) \left\{
  \begin{array}{ll}
      & (\mbox{for the case } m \neq -n) \\
    = &(m-n)\phi(K_{m+n}) - m\phi(K_m) + n\phi(K_n) \\
    \mbox{ } = &(m-n)(m+n) - m^2 + n^2 \\
    \mbox{ } = & 0 \\
    \\
      & (\mbox{for the case } m = -n) \\
    = &- m\phi(K_m) - m\phi(K_{-m}) \\
    \mbox{ } = &-m^2 - m(-m) \\
    \mbox{ } = & 0 
  \end{array}
\right. .
\]
Hence this vanishes on the commutator. The linear functional $\phi$ is nontrivial and
the commutator subalgebra $[\K_0,\K_0]$ is in the nontrivial kernel of $\phi$.
In particular, $[\K_0, \K_0]$ is not equal to $\K_0$.

To see that the commutator subalgebra of
$\mathcal{K}_0$ has codimension one, we will show that all the element of $\mathcal{K}_0$
can be obtained as the linear combination of $K_1$ and elements of $[\mathcal{K}_0, \mathcal{K}_0]$.
Let us note that 
\begin{eqnarray*}
{}  [K_1, K_{-1}] &=& - K_1 - K_{-1} \\
{}  [K_2, K_{-1}] &=& 3K_1 - 2K_{2} - K_{-1}\\
{}  [K_{-2}, K_1] &=& -3K_{-1} + 2K_{-2} + K_{1}.
\end{eqnarray*}
So $K_{-1}, K_{2}, K_{-2}$ can be obtained. For other elements in the basis, we only
need to see
\begin{eqnarray*}
{}  [K_n,K_1] &=& (n-1)K_{n+1} - nK_n + K_1 \\
{}  [K_{-n},K_{-1}] &=& -(n-1)K_{-n-1} +nK_{-n} -K_{-1},
\end{eqnarray*}
and to use mathematical induction.
\end{proof}

\begin{rmk}
In proposition 3.2 of \cite{weiner} it is claimed that
$[\mathcal{K},\mathcal{K}] = \mathcal{K}$ where $\mathcal{K}$ is the
central extension of $\mathcal{K}_0$ defined in the introduction of the present paper.
It is wrong,
as seen in lemma \ref{first}: $\mathcal{K}$, as well as $\mathcal{K}_0$,
is not perfect. In the proof of \cite{weiner}, there is a
sentence ``confronting what we have just obtained with (14), we get that ...'',
which does not
make sense. In accordance with this, the remark after proposition 3.6 and corollary
3.8 in that article should be corrected as to allow the difference by scalar.
On the other hand, what is used in corollary 3.3 is only the fact that
$\phi(C) = 0$ and the conclusion
is not changed. The main results of the paper are not at all affected.
\end{rmk}

\begin{cor}\label{1dimrep}
$H^1(\mathcal{K}_0, \mathbb{C})$ is one dimensional. In particular,
there is a unique (up to scalar) one dimensional representation of $\K_0$.
\end{cor}

Next we will determine the second cohomology group of $\K_0$.

\begin{lem}\label{basis}
The following set forms a basis of the commutator subalgebra of $\mathcal{K}_0$.
\[
  [K_n,K_1], [K_{-n},K_{-1}] \mbox{ for } n > 1, [K_{-2},K_1], [K_2,K_{-1}],
  [K_1,K_{-1}].
\]
\end{lem}
\begin{proof}
As we have seen, the commutator subalgebra is the kernel of the functional
of lemma \ref{first}. The last three elements in the set are linearly
independent and contained in the subspace spanned by $K_{-2}, K_{-1}, K_1$ and $K_2$.
The elements $[K_n,K_1]$ (respectively the elements $[K_{-n},K_{-1}]$,) contain
$K_{n+1}$ terms (respectively $K_{-(n+1)}$ terms,) hence they are independent
and form the basis of the commutator subalgebra.
\end{proof}

\begin{thm}
$H^2(\mathcal{K}_0, \mathbb{C})$ is one dimensional.
\end{thm}
\begin{proof}
Take an element $\omega$ of $Z^2(\mathcal{K}_0, \mathbb{C})$.
Let $\omega_{m,n} := \omega(K_m,K_n)$ for $m,n \in \mathbb{Z} \setminus \{0\}$
be complex numbers. From the definition of $Z^2(\mathcal{K}_0, \mathbb{C})$,
the following holds:
\begin{eqnarray}
 & & \omega_{m,n} = -\omega_{n,m} \nonumber \\
0 &=& \omega(K_l,[K_m,K_n]) + \omega(K_n,[K_l,K_m]) + \omega(K_m,[K_n,K_l]) \nonumber \\
  &=& (m-n)\omega_{l,m+n} - m\omega_{l,m} + n\omega_{l,n} \nonumber \\
  & & + (l-m)\omega_{n,l+m} - l\omega_{n,l} + m\omega_{n,m} \label{jacobi} \\
  & & + (n-l)\omega_{m,n+l} - n\omega_{m,n} + l\omega_{m,l}, \nonumber
\end{eqnarray}
where this holds also for the cases $l+m=0, m+n=0,$ or $n+l=0$ if we define $w_{k,0} = w_{0,k} = 0$
for $k \in \Z$.

Let $\alpha$ be a linear functional on the commutator subalgebra defined by
\begin{eqnarray*}
  \alpha([K_n,K_1]) &=& \omega_{n,1} \mbox{ for } n > 1\\
  \alpha([K_{-n},K_{-1}]) &=& \omega_{-n,-1} \mbox{ for } n > 1\\
  \alpha([K_{-2},K_1]) &=& \omega_{-2,1} \\
  \alpha([K_2,K_{-1}]) &=& \omega_{2,-1} \\
  \alpha([K_1,K_{-1}]) &=& \omega_{1,-1}.
\end{eqnarray*}
This definition is legitimate by lemma \ref{basis}.

If we define $\omega^\prime_{m,n} = \omega_{m,n} - \alpha([K_m,K_n])$,
there is a corresponding element $\omega^\prime$ in $Z^2(\mathcal{K}_0, \mathbb{C})$
and belongs to the same class in $Z^2/B^2(\mathcal{K}_0)$. To keep the brief notation, we
assume from the beginning the following:
\[
 \omega_{n,1} = \omega_{-n,-1} = \omega_{-2,1} = \omega_{2,-1} = \omega_{1,-1} = 0
  \mbox{ for } n > 1
\]
and we will show that $\omega_{m,n} = 0$ if $m \neq -n$.

Now we set $l = 2, m = 1, n = -1$ in (\ref{jacobi}) to get:
\[
  0 = 2\omega_{2,0} - \omega_{2,1} - \omega_{2,-1} + \omega_{-1,3} - 2\omega_{-1,2}
      + \omega_{-1,1} - 3\omega_{1,1} + \omega_{1,-1} + 2\omega_{1,2}.
\]
 From this we see that $\omega_{-1,3}$ vanishes because by assumption all the other terms
are zero. Similarly if we let $l = -2, m =1, n = 1$, we have $\omega_{1,-3} = 0$.

Furthermore, setting $l > 1, m = 1, n = -1$ we get
\begin{multline*}
  0 =  2\omega_{l,0} - \omega_{l,1} - \omega_{l,-1} + (l-1)\omega_{-1,l+1} - l\omega_{-1,l}
      + \omega_{-1,1} \\
       - (l+1)\omega_{1,l-1} + \omega_{1,-1} + l\omega_{1,l}.
\end{multline*}
This implies $\omega_{-1,l+1} = 0$ by induction for $l > 1$.
Similarly, letting $l < -1, m = 1, n = -1$ we see $\omega_{1,l-1} = 0$ for $l < -1$.

Next we use formula (\ref{jacobi}) substituting $l = 1, n = -m$ to get
\begin{multline*}
  0 =  2m\omega_{1,0} - m\omega_{1,m} - m\omega_{1,-m} + (1-m)\omega_{-m,m+1} - \omega_{-m,1}
      + m\omega_{-m,m} \\
       +(-m-1)\omega_{m,1-m} + m\omega_{m,-m} + \omega_{m,1}.
\end{multline*}
Since $\omega_{1,m} = \omega_{-1,m} = 0$, as we have seen above, and by the antisymmetry
$\omega_{-m,m} = -\omega_{m,-m}$, we have
\[
  (1-m)\omega_{-m,1+m} + (-m-1)\omega_{m,1-m} = 0.
\]
By assumption, we have $\omega_{-1,2} = 0$. By induction on $m$,
we observe $\omega_{-m,m+1} = 0$. Similarly it holds $\omega_{-m,m-1} = 0$.

Finally we fix $k \in \mathbb{N}$ and let $l = 1, n = k-m$ to get
\begin{multline*}
  0 =  (2m-k)\omega_{1,k} - m\omega_{1,m} + (k-m)\omega_{1,k-m} + (1-m)\omega_{k-m,m+1} - \omega_{k-m,1}\\
      + m\omega_{k-m,m} 
       +(k-m-1)\omega_{m,k-m+1} - (k-m)\omega_{m,k-m} + \omega_{m,1}.
\end{multline*}
By assumption, as before, the preceding equation becomes the following:
\begin{eqnarray}
  0 &=& (1-m)\omega_{k-m,m+1} + k\omega_{k-m,m} + (k-m-1)\omega_{m,k-m+1} \nonumber \\
    &=& (1-m)\omega_{(k+1)-(m+1),m+1} + k\omega_{k-m,m} + (k-m-1)\omega_{m,(k+1)-m} \label{induk}
\end{eqnarray}
If we let $k = 1$, the second term vanishes by the observation above and we see
\[
  (1-m)\omega_{1-m,m+1} -m\omega_{m,2-m} = 0
\]
Again by induction on $m$, we see $\omega_{2-m,m}$ vanishes for all $m$. Then by induction
on $k$ and using (\ref{induk}), we can conclude $\omega_{k-m,m}$ vanishes
for all $k \in \mathbb{N}, m \in \Z$. Similar argument applies for $k < 0$.

Summarizing, if we have an element in $Z^2(\mathcal{K}_0, \mathbb{C})$, we may assume
that all the off-diagonal parts vanish. Letting $l = -m-n$ in (\ref{jacobi}), we see
that there is a possibility of one (and only) dimensional second cohomology as in the
case of Virasoro algebra (see \cite{schottenloher}).
\end{proof}

This theorem shows that there is a unique central extension (up to isomorphism)
of $\K_0$. We denote the central extension by $\K$. By fixing a cocycle $\omega \in
Z^2(\K_0,\C) \setminus B^2(\K_0,\C)$ the algebra $\K$ is formally
defined as $\K_0 \oplus \C$ with the commutation relations
\[
[(x,a), (y,b)] := \left([x,y], \omega(x,y)\right) \mbox{ for } x,y \in \K_0, a,b \in \C.
\]
Equivalently, in this article and in literature, using a formal central element $C$,
one writes:
\[
[x+aC, y+bC] = [x,y]+\omega(x,y)C.
\]

\begin{prp}\label{auto}
Let us fix a real number $\lambda$.
On $\mathcal{K}$, there is a *-automorphism $\Lambda$ defined by
$\Lambda(K_n) = K_n + in\lambda C$ and $\Lambda(C) \mapsto C$.
\end{prp}
\begin{proof}
It is clear that this preserves the *-operation. Since the change by this mapping
is just an addition of a scalar multiple of the central element, this does not
change the commutator. On the other hand, as seen in lemma \ref{first},
the map $K_n \mapsto n$ vanishes on
the commutator subalgebra, hence the linear map in question preserves the commutators.
\end{proof}

\begin{prp}\label{autononextend}
The *-automorphism in Proposition \ref{auto} does not extend to the Virasoro algebra
unless $\lambda = 0$.
\end{prp}
\begin{proof}
Assume the contrary, namely that $\Lambda$ extends to $\mathrm{Vir}$. 
Since $\mathcal{K}$ has codimension one in the Virasoro algebra, we only have to
determine where $L_0$ is mapped.
The algebra $\mathrm{Vir}$ is the linear span of $K_n$'s, $C$ and $L_0$, hence
$\Lambda(L_0)$ takes the following form.
\[
  \Lambda(L_0) = \sum_{n \neq 0} a_nK_n + a_0L_0 + bC,
\]
where $a_n$'s and $b$ are complex numbers and $a_n$'s vanish except for finitely many
$n$.

On the other hand, in $\mathrm{Vir}$, we have
\[
 [K_n, L_0] = [L_n - L_0, L_0] = nL_n = nK_n + nL_0.
\]
Since in the sum of $\Lambda(L_0)$ only finitely many terms appear, let $N$ be the
largest integer with which $a_N$ does not vanish. If $N > 1$, recalling
$[K_1, L_0] = K_1 + L_0$, we have
\begin{eqnarray*}
  \Lambda([K_1, L_0]) &=& [K_1+i\lambda C, \Lambda(L_0)] \\
                      &=& \Lambda(K_1) + \Lambda(L_0),
\end{eqnarray*}
which is impossible because the second expression contains $K_{N+1}$ term but
the last expression does not. Hence $N$ must be less than 2. By the same argument
replacing $K_1$ by $K_2$, we have that $N$ must be less than 1. Similarly
replacing $K_1$ by $K_{-1}$ or $K_{-2}$, it can be shown that $\Lambda(L_0)$
must be of the form
\[
  \Lambda(L_0) = a_0L_0 + bC.
\]
We need to note that $a_0$ and $b$ must be real as $\Lambda$ is a *-automorphism.

Now let us calculate again
\begin{eqnarray*}
[\Lambda(K_1), \Lambda(L_0)] &=& [K_1 + i\lambda C, a_0L_0 + b\cdot C] \\
  &=& a_0K_1 + a_0L_0,
\end{eqnarray*}
by assumption this must be equal to
\begin{eqnarray*}
  \Lambda([K_1,L_0]) &=& \Lambda(K_1+L_0) \\
  &=& K_1 + a_0L_0 + (b + i\lambda)C,
\end{eqnarray*}
which is impossible since $b$ is real, except the case $\lambda = 0$
(and in this case $b = 0, a_0 = 1$).
\end{proof}

\begin{rmk}\label{equivauto}
When we make compositions of these automorphisms with a representation
of $\mathcal{K}$, we might obtain inequivalent representations of $\mathcal{K}$.
However these representations integrate to equivalent projective unitary
representations of the group $B_0$, since with these automorphisms
the changes of self-adjoint elements in $\mathcal{K}$ are only scalars and
the changes of their exponentials are only phases, therefore equivalent as projective
representations of $B_0$.
\end{rmk}

\section{Derived subalgebras and groups}\label{derivedsubalgebras}
\subsection{A sequence of ideals in $\K_0$}\label{commutators}
We will investigate the derived subalgebras of $\mathcal{K}_0$. The derived subalgebra
(or the commutator subalgebra) of a Lie algebra is, by definition, the subalgebra generated
by all the commutators of the given Lie algebra. 

The easiest and most important property
of the commutator subalgebra is that it is an ideal. This is clear from the definition.
If a Lie algebra is simple, then the commutator subalgebra must coincide with the Lie algebra itself.
This is the case for the Virasoro algebra.

On the other hand, the algebra $\mathcal{K}_0$ and its unique nontrivial central extension
$\mathcal{K}$ are not simple. This can be seen from lemma \ref{first}:
the commutator subalgebra (which we denote by $\mathcal{K}_0^{(1)}$) has codimension 1 in
$\mathcal{K}_0$ and it is the kernel of a homomorphism of the Lie algebra. 

Let us denote $\mathrm{Vect}(S^1)_0$ the subalgebra of $\mathrm{Vect}(S^1)$ whose element vanish at $\theta = 0$.
We remind that the commutator on $\mathrm{Vect}(S^1)$ is the following.
\begin{equation}\label{vectcomm}
  [f,g] = f g^\prime - f^\prime g.
\end{equation}
Now it is easy to see that $\mathrm{Vect}(S^1)_0$ is a subalgebra.
Let us recall that we embed $\K_0$ in $\vect_0$ by the correspondence
$K_n \mapsto i(\mathrm{exp}(in\cdot) - 1)$. 
We clarify the meaning of the homomorphism $\phi$ by considering the larger algebra $\mathrm{Vect}(S^1)_0$.

\begin{lem}
The homomorphism $\phi: K_n \mapsto -n$ on $\mathcal{K}_0$ continuously extends to
$\mathrm{Vect}(S^1)_0$ and the result is
\begin{eqnarray*}
\phi: \mathrm{Vect}(S^1)_0 &\to& \mathbb{R} \\
 f &\mapsto& f^\prime(0).
\end{eqnarray*}
\end{lem}
\begin{proof}
It is easy to see that $\phi$ and the derivative on $0$ coincide.
The latter is clearly continuous on $\mathrm{Vect}(S^1)_0$ in its smooth topology.

To see that the extension is still a homomorphism of $\mathrm{Vect}(S^1)_0$,
we only have to calculate
the derivative of $[f,g]$ on $\theta = 0$:
\begin{eqnarray*}
\frac{d}{dt}[f,g](0) &=& \left. \frac{d}{dt} \left(fg^\prime - f^\prime g\right)\right|_{t=0}\\
                     &=& \left(f^\prime g^\prime + fg^{\prime\prime}
                      - f^{\prime\prime}g - f^\prime g^\prime\right)(0) \\
                     &=& \left(f^{\prime\prime}g - fg^{\prime\prime}\right)(0) \\
                     &=& 0,
\end{eqnarray*}
since $f$ and $g$ are elements of $\mathrm{Vect}(S^1)_0$.
\end{proof}

We set $\phi_1 := \phi$ and we define similarly,
\begin{eqnarray*}
\phi_k: \mathrm{Vect}(S^1)_0 &\to& \mathbb{R} \\
 f &\mapsto& f^{(k)}(0),
\end{eqnarray*}
where $f^{(k)}$ is the $k$-th derivative of the function $f$. Again these maps
are continuous in the topology of smooth vectors.

We show the following.
\begin{lem}\label{derivative}
Let $f$ and $g$ be in $\mathrm{Vect}(S^1)_0$. Suppose $\phi_m(f) = \phi_m(g) = 0$
for $m = 1,\cdots k$.
Then $\phi_m\left([f,g]\right) = \phi_m(fg^\prime -f^\prime g) = 0$ for $m = 1, \cdots 2k+1$.
\end{lem}
\begin{proof}
First we recall the general Leibniz rule:
\[
(F \cdot G)^{(k)}(\theta)=\sum_{m=0}^k {_k}C_m F^{(m)}(\theta) G^{(k-m)}(\theta),
\]
where ${_k}C_m$ denotes the choose function $\frac{k!}{m!(k-m)!}$.
Then, in each term of the $m$-th derivatives of $[f,g] = fg^\prime - f^\prime g$
where $m \le 2k$, there appears
a factor which is a derivative $f$ or $g$ of order $m \le k$ and the term vanishes
by assumption. To consider the $(2k+1)$-th
derivative, the only nonvanishing terms are
\begin{eqnarray*}
[f,g]^{(2k+1)}(\theta) &=& {_{2k+1}}C_{k+1}f^{(k+1)}g^{(k+1)} - {_{2k+1}}C_{k}f^{(k+1)}g^{(k+1)} \\
 &=& 0.
\end{eqnarray*}
\end{proof}

\begin{prp}\label{derivation}
The subspace $\mathrm{Vect}(S^1)_k = \{f \in \mathrm{Vect}(S^1)_0: \phi_1(f) = \cdots = \phi_k(f) = 0 \}$
is an ideal of $\mathrm{Vect}(S^1)_0$ and it holds that 
\[
[\mathrm{Vect}(S^1)_k, \mathrm{Vect}(S^1)_k] \subset \mathrm{Vect}(S^1)_{2k+1}
\]
\end{prp}
\begin{proof}
The latter part follows directly from lemma \ref{derivative}.
To show that $\mathrm{Vect}(S^1)_k$ is an ideal, we only have to take
$f \in \mathrm{Vect}(S^1)_0$ and $g \in \mathrm{Vect}(S^1)_k$ and to calculate
derivatives of $[f,g]$. By the Leibniz rule above, for $m \le k$, in each
term of the $m$-th derivative of $[f,g]$ there is a factor which is a derivative
of $g$ of order less than $m$ or $f$ itself and they must vanish at $\theta = 0$
by assumption.
\end{proof}

Note that if we restrict $\phi_m$ to $\mathcal{K}_0$, it acts like $\phi_m(K_k) = i(ik)^m$.
Defining $\mathcal{K}_k = \{x \in \mathcal{K}_0: \phi_1(x) = \cdots \phi_k(x) = 0 \}$,
we can see similarly that $\{\mathcal{K}_k\}$ are ideals of $\mathcal{K}_0$ and
that $[\mathcal{K}_k, \mathcal{K}_k] \subset \mathcal{K}_{2k+1}$.

\subsection{Basis for $\mathcal{K}_k$}\label{basisfor}
Our next task is to determine the derived subalgebras of $\{\mathcal{K}_k\}$. 
For this purpose, it is appropriate to take a new basis for each $\mathcal{K}_k$.

The following observation is easy.
\begin{lem}\label{subbasis}
If $V$ is the vector space spanned by a countable basis $\{B_n\}_{n \in \mathbb{Z}}$, then
$\{B_n - B_{n+1}\}_{n \in \mathbb{Z}}$ is a linearly independent set and the vector space
spanned by them has codimension 1 in $V$.
\end{lem}

We set recursively,
\begin{eqnarray*}
M^0_n &:=& L_n - L_{n+1} \\
M^1_n &:=& M^0_n - M^0_{n+1} \\
M^{k+1}_n &:=& M^k_n - M^k_{n+1},
\end{eqnarray*}
where $\{L_n\}$ is the basis of the Witt algebra. By lemma \ref{subbasis},
we have a sequence of subspaces of $\mathrm{Witt}$. We will see that they coincide
with $\{\mathcal{K}_n$\}. For this purpose we need the combinatorial formula
in lemma \ref{subformula}.
\begin{rmk}
We use the convention that a polynomial of
degree $-1$ is 0.
\end{rmk}

\begin{lem}\label{subdeg}
If $k \ge 0$ and if $p(x)$ is a polynomial of $x$ of degree $k$, then $p(x) - p(x+1)$ is
a polynomial of degree $k-1$.
\end{lem}
\begin{proof}
We just have to consider the terms of the highest and the second highest degrees.
\end{proof}

We fix a natural number $k$. Let us define a sequence of polynomials
recursively by
\begin{eqnarray*}
p_k(x) &=& x^k, \\
p_{m-1}(x) &=& p_m(x) - p_m(x+1) \mbox{ for } 0 \le m \le k.
\end{eqnarray*}

\begin{lem}\label{subformula}
We have the explicit formulae for $-1 \le m \le k$.
\[
p_m(x) = \sum_{l=0}^{k-m} (x+l)^k(-1)^l {_{k-m}}C_l.
\]
\end{lem}
\begin{proof}
We show this lemma by induction. If $m = k$, $p_m(x) = x^k$ and
the lemma holds.

Let us assume that the formula holds for $m$.
We use the well-known combinatorial fact that if $1 < j \ \le i$ then
${_i}C_{j-1} +{_i}C_j = {_{i+1}}C_j$. Now let us calculate
\begin{eqnarray*}
p_{m-1}(x) &=& p_m(x) - p_m(x+1) \\
 &=& \sum_{l=0}^{k-m}(x+l)^k(-1)^l {_{k-m}}C_l 
     - \sum_{l=0}^{k-m}(x+1+l)^k(-1)^l {_{k-m}}C_l \\
 &=& \sum_{l=0}^{k-m}(x+l)^k(-1)^l {_{k-m}}C_l 
     - \sum_{l^\prime=1}^{k-m+1}(x+l^\prime)^k(-1)^{l^\prime-1} {_{k-m}}C_{l^\prime-1} \\
 &=& \sum_{l=0}^{k-m}(x+l)^k(-1)^l {_{k-m}}C_l 
     + \sum_{l=1}^{k-m+1}(x+l)^k(-1)^{l} {_{k-m}}C_{l-1} \\
 &=& (x+k-m+1)^k(-1)^{k-m+1} \\
 & &  + \sum_{l=1}^{k-m}(x+l)^k(-1)^l
      \left({_{k-m}}C_{l-1} + {_{k-m}}C_l\right) + x^k \\
 &=& (x+k-m+1)^k(-1)^{k-m+1} + \sum_{l=1}^{k-m}(x+l)^k(-1)^l
      {_{k-m+1}}C_l + x^k \\
 &=& \sum_{l=0}^{k-m+1} (x+l)^k(-1)^l {_{k-m}}C_l.
\end{eqnarray*}

\end{proof}

\begin{prp}\label{subpol}
For $k \ge 0$, as a polynomial of $x$, it holds
\[
\sum_{l=0}^{k+1} (x+l)^k(-1)^l {_{k+1}}C_l = 0.
\]
\end{prp}
\begin{proof}
If we put $m = -1$ in lemma \ref{subformula}, we get the left hand side
of this formula. On the other hand, by definition of $p_{-1}$ and by
lemma \ref{subdeg}, it must be a polynomial of degree $-1$, in other
words, it vanishes.
\end{proof}

We want to apply this formula to the calculation of the functionals $\phi_k$.
For this purpose we need formulae for $\{M^k_n\}$ (which are defined
at the beginning of this subsection) in terms $\{L_n\}$.

\begin{prp}\label{fullbasisformula}
It holds that
\[
M^k_n = \sum_{l=0}^{k+1}(-1)^l {_{k+1}}C_l L_{n+l}
\]
\end{prp}
\begin{proof}
Again we show this by induction. If $k=0$, then $M^0_n = L_n - L_{n+1}$ and
this case is proved.

Assume it holds $M^k_n = \sum_{l=0}^{k+1}(-1)^l {_{k+1}}C_l L_{n+l}$.
Again using the combinatorial formula ${_i}C_{j-1} +{_i}C_j = {_{i+1}}C_j$,
let us calculate
\begin{eqnarray*}
M^{k+1}_n &=& M^k_n - M^k_{n+1} \\
 &=& \sum_{l=0}^{k+1}(-1)^l {_{k+1}}C_l L_{n+l}
   - \sum_{l=0}^{k+1}(-1)^l {_{k+1}}C_l L_{n+1+l} \\
 &=& \sum_{l=0}^{k+1}(-1)^l {_{k+1}}C_l L_{n+l}
   - \sum_{l^\prime=1}^{k+2}(-1)^{l^\prime-1} {_{k+1}}C_{l^\prime-1} L_{n+l^\prime} \\
 &=& \sum_{l=0}^{k+1}(-1)^l {_{k+1}}C_l L_{n+l}
   + \sum_{l=1}^{k+2}(-1)^l {_{k+1}}C_{l-1} L_{n+l} \\
 &=& (-1)^{k+2} L_{n+k+2} 
   + \sum_{l=1}^{k+1}(-1)^l \left({_{k+1}}C_l + {_{k+1}}C_{l-1}\right) L_{n+l}
   + L_0 \\
 &=& \sum_{l=0}^{k+2}(-1)^l {_{k+2}}C_l L_{n+l}.
\end{eqnarray*}
And this is what we had to prove.
\end{proof}

\begin{cor}\label{fullbasis}
For fixed $k \ge 0$, $\{M^k_n|n \in \mathbb{Z}\}$ is a basis of
$\mathcal{K}_k$.
\end{cor}
\begin{proof}
We can extend $\phi_k$ to the Witt algebra by $\phi_k(L_n) = i(in)^k$
(for $k = 0$, $\phi_0(L_n) = i$ by definition).

Then, it is immediate that we have the following.
\begin{eqnarray*}
\mathcal{K}_0 &=& \{x \in \mathrm{Witt}: \phi_0(x) = 0 \} \\
\mathcal{K}_k &=& \{x \in \mathrm{Witt}: \phi_0(x) = \phi_1(x) = \cdots = \phi_k(x) = 0 \}.
\end{eqnarray*}
Clearly $\{\phi_k\}$ are independent and each $\mathcal{K}_{k+1}$ has
codimension 1 in $\mathcal{K}_k$.

We will prove the corollary by induction.
The set $\{M^0_n\}$ spans a subspace of $\mathrm{Witt}$ with codimension 1 by lemma
\ref{subbasis} and it is immediate to see that $\phi_0(M^0_n) = 0$.
On the other hand $\mathcal{K}_0$ is the kernel of $\phi_0$
and has codimension one in $\mathrm{Witt}$. Hence they must coincide.

Assume that $\{M^{k-1}_n\}$ is the basis of
$\mathcal{K}_{k-1}$. Then it is obvious that $M_n^k = M_n^{k-1}-M_{n+1}^{k-1} \in \mathcal{K}_{k-1}$.
Now, by proposition \ref{fullbasisformula} and proposition \ref{subformula},
we see easily that for $n \in \mathbb{Z}$
\begin{eqnarray*}
\phi_k(M^k_n) &=& \sum_{l=0}^{k+1}(-1)^l {_{k+1}}C_l \phi_k(L_{n+l}) \\
 &=& \sum_{l=0}^{k+1}(-1)^l {_{k+1}}C_l (n+l)^k \\
 &=& 0.
\end{eqnarray*}
This means that $M^k_n \in \mathcal{K}_k$.

The linear span of $\{M^k_n\}_{n \in \mathbb{Z}}$ must have codimension 1
by lemma \ref{subbasis} in $\K_{k-1}$, therefore it must coincide with
$\K_k$, since $\K_k$ has codimension 1 in $\K_{k-1}$.
\end{proof}

\subsection{Commutator subalgebras of $\mathcal{K}_k$}
Now we can completely determine all the commutator subalgebras of
$\mathcal{K}_k$. The key fact is that we can easily calculate
the commutator in the basis we have obtained in the previous section.

\begin{prp}\label{comminbasis}
Let $k \ge 0$ and $m,n \in \mathbb{Z}$.
It holds that
\[
[M^k_m, M^k_n] = (m-n)M^{2k+1}_{m+n}
\]
\end{prp}
\begin{proof}
We prove the proposition by induction. 
The case for $k = 0$ is shown as follows.
\begin{eqnarray*}
[M^0_m, M^0_n] &=& [L_m - L_{m+1}, L_n - L_{n+1}] \\
 &=& (m-n)L_{m+n} - (m+1-n)L_{m+1+n}\\
 & & - (m-n-1)L_{m+n+1} + (m-n)L_{m+1+n+1} \\
 &=& (m-n)\left(L_{m+n} - L_{m+n+1}\right) - (m-n)\left(L_{m+n+1} - L_{m+n+2}\right) \\
 &=& (m-n)(M^0_{m+n} - M^0_{m+n+1}) \\
 &=& (m-n)M^1_{m+n}
\end{eqnarray*}

Let us assume that the formula 
holds for $k$. We calculate
\begin{eqnarray*}
[M^{k+1}_m, M^{k+1}_n] &=& [M^k_m - M^k_{m+1}, M^k_n - M^k_{n+1}] \\
 &=& (m-n)M^{2k+1}_{m+n} - (m+1-n)M^{2k+1}_{m+1+n}\\
 & &  - (m-n-1)M^{2k+1}_{m+n+1} + (m-n)M^{2k+1}_{m+1+n+1} \\
 &=& (m-n)\left(\left(M^{2k+1}_{m+n} - M^{2k+1}_{m+n+1}\right)
   - \left(M^{2k+1}_{m+n+1} - M^{2k+1}_{m+n+2}\right)\right) \\
 &=& (m-n)(M^{2k+2}_{m+n} - M^{2k+2}_{m+n+1}) \\
 &=& (m-n)M^{2k+3}_{m+n}.
\end{eqnarray*}
This completes the induction.
\end{proof}

\begin{rmk}
The Witt algebra can be treated as $\K_{-1}$ in this context, in the sense that
the formula of the proposition holds for $k = -1$.
\end{rmk}

\begin{thm}\label{commalg}
It holds that $\mathcal{K}_{2k+1} = \mathcal{K}_k^{(1)}$, where $\mathcal{K}_k^{(1)}$ is the
derived subalgebra of $\mathcal{K}_k$.
\end{thm}
\begin{proof}
It is clear from corollary \ref{fullbasis} and proposition \ref{comminbasis} that the derived subalgebra
of $\mathcal{K}_k$ is included in $\mathcal{K}_{2k+1}$ and the commutators of elements
in the basis of $\K_k$ exhaust the basis of $\mathcal{K}_{2k+1}$.
\end{proof}

\subsection{The ideal structure of $\mathcal{K}_0$}
The basis obtained in the previous subsection is suitable to determine
all the ideals of $\mathcal{K}_0$. In fact, we will see that any ideal of
$\mathcal{K}_0$ must coincide with one of $\{\mathcal{K}_k\}$ or
$\ker \phi_1 \cap \ker \phi_3$.

\begin{lem}\label{largeideals}
If $\mathcal{I}$ is a nontrivial ideal of $\mathcal{K}_0$, then it includes
$\mathcal{K}_k$ for some $k$.
\end{lem}
\begin{proof}
Let $x$ be a nontrivial element of $\mathcal{I}$. It has an expansion
$x = \sum_{j=1}^N a_jM^0_{n_j}$ and we may assume $a_j \neq 0$ for all $j$.
Since $\mathcal{I}$ is an ideal of $\mathcal{K}_0$, any commutator with $x$
must be in $\mathcal{I}$ again. In particular,
\[
[M^0_{n_N}, x] = \sum_{j=1}^{N}a_j(n_N-n_j)M^1_{n_j+n_N} =
 \sum_{j=1}^{N-1} b_j M^1_{n_j+n_N},
\]
where each of $b_j = a_j(n_N-n_j),$ for $j = 1,2, \cdots N-1$, is nonzero, must be
an element of $\mathcal{I}$.

Similarly $[M^1_{n_N+n_{N-1}}, [M^0_{n_N}, x]] = \sum_{j=1}^{N-2}c_j M^1_{n_j+n_N+n_{N-1}}$
is also an element of $\mathcal{I}$. Repeating this procedure, we see that $\mathcal{I}$
contains some $M^l_m$. Then, using the commutation relation in $\mathcal{K}_l$,
we see that $\mathcal{I}$ contains $\{M^{2l+1}_n\}_{n \neq 2m}$ and
$\{M^{4l+3}_n\}_{n \in \mathbb{Z}}$. This implies that $\mathcal{I}$ includes
$\mathcal{K}_{4l+3}$.
\end{proof}

To prove the next lemma, we need to recall that $\mathcal{K}_0$ is a subalgebra
of smooth vector fields on $S^1$ and all the functionals $\{\phi_k\}_{k \in \mathbb{N}}$
have analytic interpretations as in subsection \ref{commutators}.
There, we have identified the real line with the
punctured circle, the point at infinity with the point $\theta = 0$. The algebra
$\K_0$ is realized as a subalgebra of smooth functions on the circle vanishing at
$\theta = 0$. Seen as the algebra of functions, their commutation relations are
$[x,y] = xy^\prime - x^\prime y$.

\begin{lem}\label{largecoincidence}
Let $\mathcal{I}$ be a nontrivial ideal of $\mathcal{K}_0$ and let $k$ be the
smallest number such that $\mathcal{K}_k$ is included in $\mathcal{I}$ (this
exists by lemma \ref{largeideals}). If $k \ge 4$, then $\mathcal{I} = \mathcal{K}_k$.
\end{lem}
\begin{proof}
We will prove this lemma by contradiction. Let us assume that $\mathcal{I} \neq
\mathcal{K}_k$ and that $x \in \mathcal{I} \setminus \mathcal{K}_k$.
Possible cases are (1) $x \in \K_2$ (2) $x \in \K_0 \setminus \K_1$ (3)
$x \in \K_1 \setminus \K_2$. We treat these cases in this order.

If $x \in \mathcal{K}_2$, then there is $l$ such that $2 \le l < k$ and
$x \in \mathcal{K}_l \setminus \mathcal{K}_{l+1}$. Let us take an element $y$ from
$\mathcal{K}_1 \setminus \mathcal{K}_2$. Then, since $\mathcal{K}_l$ is an
ideal of $\mathcal{K}_0$ by the remark after proposition \ref{derivation},
we see $[x,y] \in \mathcal{K}_l$ and we calculate the derivatives at
$\theta = 0$. By the assumption on $x$ and $y$, the derivatives vanish up to
certain orders and we have the following:
\begin{align*}
[x,y]^{(l+1)}(0) &= \sum_{k=0}^{l+1} {_{l+1}}C_k \left(y^{(k)}(0) x^{(l+1-k+1)}(0)
 - y^{(k+1)}(0) x^{(l+1-k)}(0)\right) \\
  &= 0,
\end{align*}
\begin{align*}
[x,y]^{(l+2)}(0) &= \sum_{k=0}^{l+2} {_{l+2}}C_k \left(y^{(k)}(0) x^{(l+2-k+1)}(0)
 - y^{(k+1)}(0) x^{(l+2-k)}(0)\right) \\
 &= ({_{l+2}}C_2 -  {_{l+2}}C_1)y^{(2)}(0)x^{(l+1)}(0) \\
 &= \frac{(l+2)(l-1)}{2}y^{(2)}(0)x^{(l+1)}(0).
\end{align*}
The latter cannot be zero by assumption and the fact $2 \le l$. This means
$[x,y]$ is in $\mathcal{K}_{l+1} \setminus \mathcal{K}_{l+2}$. Repeating
this procedure, we obtain an element of $\mathcal{I}$ in $\mathcal{K}_{k-1} \setminus \mathcal{K}_{k}$.
Therefore $\mathcal{I}$ contains $\K_{k-1}$ because by definition $\mathcal{I}$ contains
$\K_k$ and $\K_{k-1}$ has codimension 1 in $\K_{k-1}$.
But this contradicts the definition of $k$ and we see that $x \in \mathcal{K}_2$ is impossible.

Next we $x \in \mathcal{K}_0 \setminus \mathcal{K}_1$. Then we can expand $x = a_0M^0_0 + a_1M^1_0 + y$ 
(here we use same symbols as before to save the number of characters)
where $y \in \mathcal{K}_2$, hence $a_0$ is nonzero.
If $a_1 \neq 0$ we have $[M^0_0, x] = a_1M^1_1 + [M^0_0,y]$.
If $a_1 = 0$ we have $[M^0_1, x] = a_0M^1_1 + [M^0_1, y]$.
Therefore at least one of these is in $\mathcal{K}_1 \setminus \mathcal{K}_2$ and
we may assume that $x \in \K_1 \setminus \K_2$.

Let us assume that $x \in \K_1 \setminus \K_2$. Here we consider the following
two cases, namely (3-1) $\phi_3(x) \neq 0$ (3-2) $\phi_3(x) = 0$.
If $\phi_3(x) \neq 0$ and $y \in \mathcal{K}_0 \setminus \mathcal{K}_1$, then
we see that $[x,y] + y^\prime (0)x \in \mathcal{K}_2 \setminus \mathcal{K}_3$
(and this element is clearly in $\mathcal{I}$). In fact, by a direct calculation
or by the Leibniz rule, we see
\begin{eqnarray*}
\left([x,y] + y^\prime(0) x\right)^{(2)}(0) &=& -y^\prime(0)x^{(2)}(0) + y^\prime(0)x^{(2)}(0) \\
 &=& 0, \\
\left([x,y] + y^\prime(0) x\right)^{(3)}(0) &=& -y^\prime(0)x^{(2)}(0) + y^\prime(0)x^{(2)}(0) \\
 &=& -2y^\prime(0)x^{(3)}(0) + y^\prime(0) x^{(3)}(0) \\
 &=& -y^\prime(0)x^{(3)}(0).
\end{eqnarray*}
This implies that there is an element of $\mathcal{I}$ in
$\mathcal{K}_2 \setminus \mathcal{K}_3$. By repeating the argument in the paragraph
for the case $x \in \K_2$, we see again a contradiction. Hence we must have $\phi_3(x) = 0$.

By the calculation above, this time $[x,y] + y^\prime(0)x \in \mathcal{K}_3$, but
using $\phi_3(x) = 0$ we see
\begin{eqnarray*}
\left([x,y] + y^\prime(0) x\right)^{(4)}(0) &=& 2y^{(3)}(0)x^{(2)}(0)-3y^\prime(0)x^{(4)}(0) + y^\prime(0)x^{(4)}(0) \\
 &=& 2y^{(3)}(0)x^{(2)}(0)-2y^\prime(0)x^{(4)}(0).
\end{eqnarray*}
Hence with an appropriate element $y$ this does not vanish. That means $[x,y] + y^\prime(0)x$ is an element
of $\mathcal{K}_3 \setminus \mathcal{K}_4$. By the same argument as in the case of $x \in \K_2$,
we see that this contradicts the definition of $k$ and this completes the proof.
\end{proof}

We state now the final result of this subsection.
\begin{thm}\label{idealstructure}
If $\mathcal{I}$ is an ideal of $\mathcal{K}_0$, then the possibilities are
\begin{itemize}
\item $\mathcal{I} = \{0\}$
\item $\mathcal{I} = \mathcal{K}_k$ for some $k \ge 0$
\item $\mathcal{I} = \ker \phi_1 \cap \ker \phi_3$.
\end{itemize}
\end{thm}
\begin{proof}
As before, we can define a number $k \ge 0$ as the smallest number
such that $\mathcal{K}_k$ is included in $\mathcal{I}$.

If $k=0$ or $k=1$, then there is nothing to do because the former case
means $\mathcal{I} = \mathcal{K}_0$ and in the latter case $\mathcal{K}_1$
has already codimension 1 and $\mathcal{I}$ must coincide with it.

Next we consider the case $k=2$. Since $\mathcal{K}_2$ has codimension 2 in $\mathcal{K}_0$,
it holds $\mathcal{I} = \mathcal{K}_2$ or $\mathcal{I}$ has an extra element.
But the latter case cannot happen because if $x \in \mathcal{I} \setminus \mathcal{K}_2$
we can expand $x = a_0M^0_0 + a_1M^1_0 + y$ (the same symbols again, but the coefficients
of a different element) where $y \in \mathcal{K}_2$ and $a_0 \neq 0$
(since otherwise $x \in \mathcal{K}_1$ and contradicts the assumption that $k=2$).
If $a_1\neq 0$ then
$[M^0_0, x] = a_1M^1_1 + [M^0_0,y] \in \mathcal{K}_1 \setminus \mathcal{K}_2$.
If $a_1=0$ then $[M^0_1, x] = a_0M^1_1 + [M^0_1, y] \in \mathcal{K}_1 \setminus \mathcal{K}_2$.
In both cases they contradict the assumption $k = 2$.

Let us assume $k=3$ and $\mathcal{I} \neq \mathcal{K}_3$. We can take
an element $x \in \mathcal{I} \setminus \mathcal{K}_3$ and expand it
as
\[
x = a_0M^0_0 + a_1M^1_0 + a_2M^2_0 + y
\]
(same symbols again to different coefficients)
where $y \in \mathcal{K}_3$.
By straightforward calculations we see that:
\begin{align*}
[M^0_0, x] &= a_1M^1_1 + 2a_2M^2_1 + [M^0_0, y] \\
[M^0_1, x] &= (a_0+a_1)M^1_1 + a_2(M^2_1+M^2_2) [M^0_1,y] \\
[M^0_1, [M^0_0,x]] &= a_1M^1_3 + 4a_2M^2_3 + [M^0_1, [M^0_0, y]] \\
[M^0_1, [M^0_1, x]] &= (a_0+a_1)M^1_3 + a_2(M^2_3 - 3M^2_4).
\end{align*}
We note that all these elements are in $\mathcal{I}$ since it is an ideal.
By comparing the first and third equations, we see
\begin{align*}
[M^0_0, x] - [M^0_1, [M^0_0,x]] = a_1(M^2_1 + M^2_2) + 2a_2(M^2_1 - 2M^2_3) + z \\
 = 2(a_1-a_2)M^2_3 + a_1(M^3_1 + 2M^3_2) + 2a_2(M^3_1 + M^3_2) + z,
\end{align*}
where $z$ is the sum of commutators of $y$ and hence again in $\mathcal{K}_3$.
Now it is easy to see that this element is in $\mathcal{K}_3$ if and only if
$a_1=a_2$. And this must be in $\mathcal{K}_3$, since otherwise it is in
$\mathcal{K}_2 \setminus \mathcal{K}_3$ and contradicts the assumption that
$k = 3$. Therefore we have $a_1 = a_2$.

Next we consider the difference of the second and fourth equations with
$a_1=a_2$ above and we get
\begin{eqnarray*}
[M^0_1, x] - [M^0_1, [M^0_1,x]] &=& (a_0+a_1)(M^2_1 + M^2_2) + \\
 & & a_1(M^2_1 + M^2_2 - M^2_3 -3M^2_4) + z^\prime \\
 &=& a_0(M^2_1+M^2_2) \\
 & & + a_1(2M^2_1+2M^2_2-M^2_3-3M^2_4) + z^\prime \\
 &=& a_0(M^2_1+M^2_2)+a_1(2M^3_1+4M^3_2+3M^3_3) + z^\prime,
\end{eqnarray*}
where $z^\prime$ is again an element of $\mathcal{K}_3$. As before it is in
$\mathcal{I}$. By the assumption $k = 3$ it is contained in $\mathcal{K}_3$, therefore $a_0 = 0$.
This indicates that an extra element of $\mathcal{I}$ must have the form
\[
x = a_1 (M^1_0 +  M^2_0)
\]
and it is immediate to see this is in $\ker \phi_1 \cap \ker \phi_3$.
Since $\K_3$ has codimension 1 in this intersection, $\mathcal{I}$ must be
equal to $\ker \phi_1 \cap \ker\phi_3$.

By calculating derivatives, we can see that $\ker \phi_1 \cap \ker \phi_3$ is
surely an ideal of $\mathrm{Vect}(S^1)_0$ and it is also the case even when
restricted to $\mathcal{K}_0$.

The case $k \ge 4$  is already done in lemma \ref{largecoincidence}.
\end{proof}

\subsection{The derived subgroup of $B_0$}\label{thederived}
As mentioned in the introduction, $\mathrm{Diff}(S^1)$ is the group of smooth,
orientation preserving diffeomorphisms of $S^1$. The group $B_0$
is the subgroup of $\mathrm{Diff}(S^1)$ whose elements fix the point $\theta = 0$.
Identifying $S^1$ and $\R/2\pi\Z$, we can think of an element of $B_0$ as a smooth function
$g$ on $\mathbb{R}$, satisfying $g(\theta+2\pi) = g(\theta)+2\pi$, $g(0) = 0$ and $g^\prime(\theta) > 0$.
The last condition comes from the fact that $g$ has a smooth inverse. On the other hand,
a function on $\mathbb{R}$ with the conditions above can be considered as an element of $B_0$.
And it is easy to see that the composition operation of the group coincides with the
composition of functions. In what follows we identify the group $B_0$ with the set of smooth
functions with these conditions.

Under this identification, Lie algebra $\vect$ of $B_0$ is seen as the space of
smooth functions $f$ such that $f(0) = 0$ and $f(\theta + 2\pi) = f(\theta)$. 

\begin{prp}
$B_1 := \{g \in B_0: g^\prime(0) = 1\}$ is a subgroup of $B_0$.
\end{prp}
\begin{proof}
By a simple calculation.
\end{proof}

\begin{prp}
The derived group $[B_0,B_0]$ is included in $B_1$.
\end{prp}
\begin{proof}
Take elements $g,h$ from $B_0$. It holds that
\begin{eqnarray*}
\frac{d}{d\theta}[g,h](0) &=& \frac{d}{d\theta}\left(g\circ h\circ g^{-1}\circ h^{-1} \right)(0) \\
 &=&
 g^\prime(h(g^{-1}(h^{-1}(0))))\times h^\prime(g^{-1}(h^{-1}(0))) \\
 & & \times (g^{-1})^\prime(h^{-1}(0)) \times (h^{-1})^\prime(0) \\
 &=& g^\prime(0) \times h^\prime(0) \times (g^{-1})^\prime(0) \times (h^{-1})^\prime(0) \\
 &=& 1,
\end{eqnarray*}
where the last equality holds since the derivative of the inverse function on the corresponding
point is the inverse number.
\end{proof}

We need the following well-known result \cite{thurston} \cite{mather} \cite{epstein}.
\begin{thm}
The group $\mathrm{Diff}(\mathbb{R})_c$ is simple, where $\mathrm{Diff}(\mathbb{R})_c$ is the
group of smooth
orientation-preserving diffeomorphisms of $\mathbb{R}$ whose supports are compact.
\end{thm}
Here, a support of a diffeomorphism means the closure of the set on which the given
diffeomorphisms is not equal to the identity map.

\begin{cor}\label{cut}
Let $B_c$ be the subgroup of $B_0$ whose elements have supports
not containing $\theta = 0$. Then $B_c$ is simple.
\end{cor}
\begin{proof}
There is a smooth diffeomorphism between $\mathbb{R}$ and $S^1 \setminus \{0\}$,
for example, the stereographic projection.
This diffeomorphism induces an isomorphism between $\mathrm{Diff}(\mathbb{R})_c$
and $B_c$.
\end{proof}

The following is a result similar to the fact $[\K_0,\K_0] = K_1$ which we have
proved in theorem \ref{commalg}.

\begin{thm}
$[B_0,B_0]$ is dense in $B_1$.
\end{thm}
\begin{proof}
By corollary \ref{cut}, $B_0$ has a simple subgroup $B_c$.
The simplicity of $B_c$ implies $[B_0,B_0]$ includes $[B_c,B_c] = B_c$,
since any commutator subgroup is normal. Hence we can freely use compactly supported
diffeomorphisms.

Let $g$ be an element of $B_1$. By the observation above, there is an element $h$ of $[B_0,B_0]$
such that $g \circ h$ has compact support around $0$. In other words, we may assume that $g$
has a compact support around $0$ and we only have to approximate $g$ with elements in $[B_0, B_0]$.

By the stereographic projection in corollary \ref{cut},
we can consider $g$ as a diffeomorphism of $\mathbb{R}$.
It is well-known that dilations of $\mathbb{R}$ are mapped by this isomorphism
to elements of $B_0$.
Let $\delta_t$ be the dilation by $t$.
For $x \in \mathbb{R}$, it holds
\[
\delta_t^{-1} \circ g^{-1} \circ \delta_t(x) = \frac{1}{t}g^{-1}(tx).
\]
By assumption $g^\prime(0) = 1$. It easy to see that for $t \to 0$ the functions $\frac{1}{t}g^{-1}(tx)$,
its first derivative and higher-order derivatives converge to $x$, $1$ and $0$ respectively, uniformly on each
compact set of $\mathbb{R}$. This means $\frac{1}{t}g^{-1}(tx)$ approximates the identity map
around $x = 0$.

Let $\epsilon$ be a positive number. Let $\gamma$ be a smooth positive function on $\mathbb{R}$ such that
it is $1$ on $[-\epsilon, \epsilon]$ and $0$ on $x \le -2 \epsilon$ or $x \ge 2\epsilon$.
And let us consider the following functions parametrized by $t$.
\[
 h_t(x) = x + \left(\frac{1}{t}g^{-1}(tx) - x\right)\gamma(x).
\]
It is easy to see that $h_t$'s are smooth, $h_t(0) = 0$, $h_t$'s are equal to 
$x$ outside a compact set and if $t$ is sufficiently small then each of $h_t$ has
the first derivative which is strictly larger than $0$.
Hence we can consider $h_t$ as a diffeomorphism of $\mathbb{R}$ with a compact support.
 From the observation above it is clear that $h_t$ and its derivatives converge
to $x$, $1$, and $0$ uniformly on $\mathbb{R}$, namely $h_t$ converge to the
identity element in the smooth topology.

An important fact is that $h_t$ is equal to $\frac{1}{t}g^{-1}(t\cdot)$ on $[-\epsilon, \epsilon]$.
The map $\delta^{-1}_t \circ g \circ \delta_t \circ h_t$ has a compact support which
does not contain $0$, hence it corresponds to an element of $B_c$. We denote it by $f_t$.

Now it is evident that $(g \circ \delta^{-1}_t \circ g^{-1} \circ \delta_t) \circ f_t
= g \circ (\delta^{-1}_t \circ g^{-1} \circ \delta_t \circ f_t)$ is in
$[B_0,B_0]$ because it is a product of a commutator and a diffeomorphism with compact support.
It is equal to $g \circ h_t$ which converges to $g$ with all its derivatives.
This shows $[B_0,B_0]$ is dense in $B_1$.
\end{proof}

\begin{rmk}
The Lie group $\mathrm{Diff}(S^1)$ is simple \cite{thurston} \cite{mather} \cite{epstein}, but
the Lie algebra $\mathrm{Vect}(S^1)$ is not simple. This is easy to see: for example,
we only have to consider the subalgebra of vector fields with compact supports in
some fixed proper subinterval of $S^1$. By the commutation relation (\ref{vectcomm})
this subalgebra is an ideal. This is closed in the smooth topology, hence $\mathrm{Vect}(S^1)$
is not even topologically simple.

On the other hand, the Witt algebra is simple. This can be seen by observing
that the linear map $[L_0, \cdot]$ is diagonalized on the standard basis of $\witt$
with no degeneration and
that we can raise or lower the elements by commutating with $L_n$ or $L_{-n}$. From this
it is easy to see that any ideal containing nontrivial element must contain $\witt$.
\end{rmk}

\section{The automorphism group of $\mathcal{K}$}\label{theautomorphism}
In this section we will completely determine the *-automorphism group of $\mathcal{K}$,
the unique central extension of $\K_0$ defined in section \ref{cohomology}.
However, this group is not necessarily a natural object. As we have seen in the introduction,
the algebra $\mathcal{K}_0$ is a subalgebra of $\vect_0$, the Lie algebra of vector
fields on $S^1$ which vanish at $\theta = 0$. On this algebra of vector fields the
stabilizer subgroup $B_0$ of $\theta = 0$ of $\mathrm{Diff}(S^1)$ acts as automorphisms,
but when we restrict these actions to $\mathcal{K}_0$, it does not necessarily globally
stabilize $\mathcal{K}_0$. In fact, the group of *-automorphisms turns out to be very small.
The situation is similar for the Virasoro algebra \cite{zhao}.

We will study this problem only for the interest of representation theory. Many things
are known about the representation theory of Virasoro algebra. In particular, all the
irreducible unitary highest weight representations are completely classified
\cite{KR}.
But for the algebra $\mathcal{K}$ the situation is different. Of course we can restrict
any unitary representation of the Virasoro algebra to $\mathcal{K}$ to obtain a unitary
representation of $\mathcal{K}$. But it is not known if there are other unitary
representations which are not localized at the point at infinity.

On the other hand, if we make a composition of a (known) unitary representation with an
endomorphism of $\mathcal{K}$ then we obtain a (possibly new) unitary representation.
The result will show that this method is not productive and, in fact, all the representations
obtained by this method are already known.

The algebra $\mathcal{K}$ has a natural decomposition
$\mathcal{K} = \mathcal{K}_+ \oplus \mathcal{K}_- \oplus \mathbb{C}C$ where
$\mathcal{K}_+ = \mathrm{span}\{K_n: n \ge 1\}$,
$\mathcal{K}_- = \mathrm{span}\{K_n: n \le -1 \}$.
Each of these direct summands is a subalgebra and it holds $\mathcal{K}_+^* = \mathcal{K}_-$.

\begin{lem}\label{proportional}
Let $K$ and $K^\prime$ be elements of $\mathcal{K}$.
We expand them in the standard basis:
\begin{eqnarray*}
K &=& a_0C + a_{n_1} K_{n_1} + a_{n_2} K_{n_2} + \cdots + a_{n_k} K_{n_k}, \\
K^\prime &=& b_0C + b_{m_1} K_{m_1} + b_{m_2} K_{m_2} + \cdots + b_{m_l} K_{m_l}.
\end{eqnarray*}
We assume here that all $a_{n_i}$ and $b_{m_j}$ but $a_0$ and $b_0$ are not zero and
that $n_1 < n_2 < \cdots < n_k$ and $m_1 < m_2 < \cdots < m_l$.
Suppose the expansion of $[K,K^\prime]$ in the standard basis does not contain terms $K_i$
where $i > \max\{n_k, m_l\}$. If we decompose $K = K_+ + K_- + a_0C$ and
$K^\prime_+ + K^\prime_- + b_0C$ according
to the decomposition $\mathcal{K} = \mathcal{K}_+ \oplus \mathcal{K}_- \oplus C\mathbb{C}$,
then $K_+$ and $K^\prime_+$ are proportional.
\end{lem}
\begin{proof}
We take a look of the commutation relations (\ref{kcomm}) of $\mathcal{K}$.
It is easy to see that in $[K_m, K_n]$ the term with index higher than $m$ and $n$
appears if and only if $m$ and $n$ are positive. And in such a case, the term
$K_{m+n}$ appears if $m \neq n$.

We may assume $n_k$ and $m_l$ are positive, since otherwise the statement would
be trivial.

 From the observation above, we see that $n_k$ must be equal to $m_l$. Otherwise,
the term $K_{n_k+m_l}$ (which is larger than $\max\{n_k,m_l\}$) appears in $[K,K^\prime]$
and cannot be cancelled, but this contradicts
the assumption that there is no term with index higher than $n_k$ and $m_l$ in the
commutator.

Now $K$ and $K^\prime$ have the following form:
\begin{eqnarray*}
K &=& a_{n_1} K_{n_1} + a_{n_2} K_{n_2} + \cdots + a_{n_{k-1}} K_{n_{k-1}} + a_{n_k} K_{n_k}, \\
K^\prime &=& b_{m_1} K_{m_1} + b_{m_2} K_{m_2} + \cdots + b_{m_{l-1}} K_{m_{l-1}} + b_{m_l} K_{n_k}.
\end{eqnarray*}
In the commutator $[K,K^\prime]$, the terms with the highest indices are now $K_{n_k+n_{k-1}}$ and
$K_{n_k+m_{l-1}}$ which appear from the commutators of $K_{n_k}$ and $K_{n_{k-1}}$ or
$K_{m_{l-1}}$. If one of $n_{k-1}$ and $m_{l-1}$ is still positive, then
again by the assumption, the highest term in the commutator must be cancelled.
This implies that again $n_{k-1} = m_{l-1}$ and $a_{n_k}b_{m_{l-1}} = b_{m_l}a_{n_{k-1}}$.
This means that the last two terms of $K$ and $K^\prime$ are proportional.

Next steps go similarly: we know the last two terms are proportional and their
commutator vanishes. Again by considering the terms with highest indices
which appear from the commutator
$[K,K^\prime]$, we see also that the last three terms are proportional.
Continuing this procedure,
we can see that all the positive part of $K$ and $K^\prime$ must be proportional.
\end{proof}

Note that with a completely analogous proof we can show a similar lemma for the negative
parts.

\begin{lem}\label{k1prop}
If $\rho$ is a *-endomorphism of $\mathcal{K}$, then there is an element $K$ of $\mathcal{K}_+$
and $\lambda, \mu, \nu \in \mathbb{C}$ such that $\rho(K_1)$ takes the form
\[
 \rho(K_1) = \lambda K + \mu K^* + \nu C.
\]
\end{lem}
\begin{proof}
Since $\rho$ is a *-endomorphism, it holds that $\rho(K_{-1}) = \rho(K_1)^*$ and
from (\ref{kcomm})
\[
[\rho(K_1), \rho(K_1)^*] = - \rho(K_1) - \rho(K_1)^*. 
\]

We can apply lemma \ref{proportional} to see that the positive part of
$\rho(K_1)$ is proportional to the positive part of $\rho(K_1)^*$. This is the
statement of the lemma.
\end{proof}

With an analogous argument we have the following:
\begin{lem}\label{k2prop}
If $\rho$ is a *-endomorphism of $\mathcal{K}$, then there is an element $K^*$ of $\mathcal{K}_+$
and $\lambda^\prime, \mu^\prime, \nu^\prime \in \mathbb{C}$ such that $\rho(K_2)$ takes the form:
\[
 \rho(K_2) = \lambda^\prime K^\prime + \mu^\prime K^{\prime*} + \nu^\prime C.
\]
\end{lem}

By a direct calculation we see that the map $\tau$ defined by
\[
\tau(K_n) = -K_{-n},  \tau(C) = -C
\]
is a *-automorphism of $\mathcal{K}$ (it extends also to the Virasoro algebra).

It is also immediate that $C$ is the unique central element up to a scalar.
This means that any automorphism must map $\mathbb{C}C$ to $\mathbb{C}C$.

\begin{lem}\label{autopositive}
If $\rho$ is a *-automorphism of $\mathcal{K}$, then there are two possibilities.
\begin{enumerate}
\item There are elements $K,K^\prime$ of
$\mathcal{K}_+$ and $\nu, \nu^\prime \in \mathbb{C}$ such that
$\rho(K_1) = K + \nu C$ and $\rho(K_2) = K^\prime + \nu^\prime C$.
\item There are elements $K,K^\prime$ of
$\mathcal{K}_-$ and $\nu, \nu^\prime \in \mathbb{C}$ such that
$\rho(K_1) = K + \nu C$ and $\rho(K_2) = K^\prime + \nu^\prime C$.
\end{enumerate}
\end{lem}
\begin{proof}
By lemma \ref{k1prop}, $\rho(K_1)$ takes the form $\rho(K_1) = \lambda K + \mu K + \nu C$ where
$K \in \mathcal{K}_+$ and $\lambda,\mu, \nu \in \mathbb{C}$.
By lemma \ref{k2prop} we have that
$\rho(K_2) = \lambda^\prime K^\prime + \mu^\prime K^{\prime*} + \nu^\prime$.
Let us recall that the following commutation relation holds.
\[
[\rho(K_2), \rho(K_1)^*] = 3\rho(K_1) - 2\rho(K_2) - \rho(K_1)^*.
\]
Note that $\rho(K_1)^* = \overline{\mu}K + \overline{\lambda}K^* + \overline{\nu}C$.

By considering the composition
with $\tau$, we may assume that $\lambda \neq 0$ ($\lambda = \mu = 0$ is impossible because
it would mean that $K_1$ is mapped to a central element and $\rho$ would not
be an automorphism). We show that $\mu = 0$. If not, applying lemma \ref{proportional} we see
that $K^\prime$ must be proportional to $K$. But this is impossible because
we would have
\begin{eqnarray*}
\rho(K_1) &=& \lambda K + \mu K^* + \nu C, \\
\rho(K_1)^* &=& \overline{\mu} K + \overline{\lambda} K^* + \overline{\nu}C, \\
\rho(K_2) &=& \lambda^\prime K + \mu^\prime K^* + \nu^\prime C, \\
\rho(K_2)^* &=& \overline{\mu^\prime} K + \overline{\lambda^\prime} K^* + \nu^\prime C,
\end{eqnarray*}
which are linearly dependent. The map $\rho$ is an automorphism and this is
a contradiction. Similarly we have $\mu^\prime = 0$ applying lemma \ref{proportional}
to the negative parts of $\rho(K_1)^*$ and $\rho(K_2)$. This concludes the lemma.
\end{proof}

Now we can determine all the elements of the *-automorphism group of $\mathcal{K}$.
Recall there is a family $\Lambda$ of *-automorphisms parametrized by $\lambda \in \R$
defined in proposition \ref{auto}.

\begin{thm}\label{classauto}
If $\rho$ is a *-automorphism of $\mathcal{K}$, then $\rho = \Lambda$ for some
$\lambda \in \mathbb{R}$ or $\rho = \Lambda \circ \tau$.
\end{thm}
\begin{proof}
By lemma \ref{autopositive} and possibly a composition with $\tau$, we may assume that
$\rho(K_1) = K + \nu C$ and $\rho(K_2) = K^\prime + \nu^\prime C$ where $K$ and
$K^\prime$ are in $\mathcal{K}_+$.

Let us expand $K$ and $K^\prime$ in the standard basis of $\mathcal{K}$,
\[
K = \sum_{i=1}^N a_iK_i, K^\prime = \sum_{j=1}^M b_jK_j,
\]
and assume $a_N \neq 0 \neq b_M$.

Since $\rho$ is a *-automorphism, it must hold that
\begin{align}
[\rho(K_1), \rho(K_1)^*] &= - \rho(K_1) - \rho(K_1)^*, \label{comm11}\\
[\rho(K_2), \rho(K_2)^*] &= - 2 \rho(K_2) - 2 \rho(K_2)^* + \frac{C}{2}, \label{comm22}\\
[\rho(K_2), \rho(K_1)^*] &= 3\rho(K_1) - 2\rho(K_2) - \rho(K_1)^* \label{comm21}.
\end{align}

Considering the terms $K_{N}$ in the first equation,
we see that $\sum_{i=1}^N a_i = \frac{1}{N}$.
Similarly, considering the terms $K_{M}$ in the second equation,
we obtain $\sum_{j=1}^M b_j = \frac{2}{M}$.
On the other hand, by comparing the terms $K_{-N}$ in the third equation, it turns out that
$-N\overline{a_N}\sum_{j=1}^M b_j = - \overline{a_N}$.
Since we have the assumption that $a_N$ is not zero, this implies that $2N = M$.

The subalgebra $\mathcal{K}_+$ is generated by $K_1$ and $K_2$ with the recursive formula
\[
K_{n+1} = \frac{1}{n-1}\left([K_n,K_1] + nK_n - K_1\right).
\]
 From this formula we see by induction that the term with the highest index
of $\rho(K_k)$ is $K_{kN}$.
If $N$ was larger than 1, these terms would not span all of $\mathcal{K}_+$ and $\rho$
could not be surjective. Thus $N$ must be $1$.

Again, by equation (\ref{comm11}) and by a direct calculation, we obtain $a_1 = 1$,
namely:
\[
\rho(K_1) = K_1 + \nu_1 C,
\]
where $\nu$ is a pure imaginary number. Similarly
we have two solutions for equation (\ref{comm22}):
\[
\rho(K_2) = \left\{
\begin{array}{l}
K_2 + \nu_2 C, \\
-\frac{1}{3}K_1 + \frac{4}{3}K_2 + \nu_2 C.
\end{array}
\right.
\]
The second solution does not satisfy equation (\ref{comm21}).
Then again by (\ref{comm21}) we see $2\nu_1 = \nu_2$.

We have seen in proposition \ref{auto} that this $\rho$ can surely be extended
to a *-automorphism of $\mathcal{K}$.
Since $K_1$ and $K_2$ are the generators of $\mathcal{K}$ as a *-Lie algebra,
this determines $\rho$ uniquely.
\end{proof}

\begin{cor}
$Aut(\mathcal{K}) \cong \mathbb{R} \rtimes \mathbb{Z}_2$.
\end{cor}

\begin{rmk}
It is also possible to determine the automorphism group of the Virasoro algebra
\cite{zhao}:
it is generated by the extension of $\tau$ and one-parameter subgroup of rotation:
\begin{eqnarray*}
\rho_t(L_n) &=& e^{itn}L_n,\\
\rho_t(C) &=& C.
\end{eqnarray*}
It is again isomorphic to $\mathbb{R} \rtimes \mathbb{Z}_2$, but the action
of the $\mathbb{R}$ part is different.
\end{rmk}

\section{Generalized Verma modules}\label{generalizedverma}
As we have seen in section \ref{cohomology}, $\mathcal{K}_0$ has the unique (up to isomorphism)
central extension which is a subalgebra of $\mathrm{Vir}$. We denote it by $\mathcal{K}$.
This section is an attempt to construct a family of unitary representations of
$\mathcal{K}$.

We are going to construct modules $V_{h+ih^\prime, c, \lambda}$ parametrized by three
complex numbers $h+ih^\prime, c, \lambda$, where $h, h^\prime \in \mathbb{R}$
and $c, \lambda \in \mathbb{C}$.
Every module has a ``lowest weight vector''  which satisfies
$K_n v = (h+ih^\prime+n\lambda)v$ for $n \ge 1$ and $Cv = cv$. If we restrict to the case
$\lambda = 0$, this module reduces to the restriction of the Virasoro module to $\mathcal{K}$.

Recall that $\K$ is a *-Lie algebra. A sesquilinear form $\langle \cdot,\cdot\rangle$
on a module $V$ is said to be contravariant if for any $v,w \in V$ and $x \in \K$ it holds
$\langle xv,w\rangle = \langle v,x^*w\rangle$. In addition if this sesquilinear
form is positive definite, then the representation of $\K$ on $V$ is said to be unitary.

It turns out that for any set of values of $h, h^\prime, c, \lambda$ we can construct a
corresponding module. In addition, if $c$ is real, there exists a contravariant sesquilinear
form on the module. Then we arrive at natural problems, for example,
when the contravariant form is unitary, when the representation of $\mathcal{K}$
integrates to the (projective unitary)
representation of $B_0$ and when these representations are inequivalent, etc.
These problems will be addressed in further publications of the author.

Here we make some remarks. It is easy to see that these modules are inequivalent as
representations of the Lie algebra $\mathcal{K}$, however, as we saw in the remark
\ref{equivauto}
(after proposition \ref{autononextend}), the imaginary part of $\lambda$ does not make
difference for the corresponding projective representation of the group $B_0$.
In addition, in \cite{weiner}
it has been proved that there are modules which integrate to equivalent projective
representations of the group for some different values of $h$. Furthermore, as we will see
in section \ref{someunitary}, there exist true (non projective) representation of $B_0$
whose naturally corresponding representations of $\mathcal{K}$ are not lowest weight modules.
In the case of $\mathrm{Diff}(S^1)$ there is a one-to-one correspondence between
irreducible unitary positive energy projective representations of the group and
irreducible lowest weight unitary representations of the Virasoro algebra.
But for $B_0$ and $\mathcal{K}$ we cannot expect such a correspondence.

\subsection{General construction of modules}\label{generalconstruction}
We start with general notions.
Let $\mathcal{L}_0$ be a Lie algebra, $U(\mathcal{L}_0)$ the universal enveloping
algebra of $\mathcal{L}_0$, $\psi_0$ a nontrivial linear functional
on $\mathcal{L}_0$ which vanishes on the commutator
subalgebra $[\mathcal{L}_0, \mathcal{L}_0]$. In particular, we assume that $\mathcal{L}_0$
is not semisimple (otherwise $\psi_0$ would be trivial). Later
$\mathcal{L}_0$ will be a upper-triangular subalgebra of a Lie algebra.
\begin{lem}\label{mult}
The linear functional $\psi_0$ extends to a homomorphism of the universal algebra
$U(\mathcal{L}_0)$.
\end{lem}
\begin{proof}
Clearly $\psi_0$ extends to a homomorphism of the tensor algebra of $\mathcal{L}_0$.
Now we only have to recall that $U(\mathcal{L}_0)$ is the quotient algebra by the two-sided
ideal generated by elements of the form $a\otimes b - b\otimes a - [a,b]$ where
$a,b \in \mathcal{L}_0$. By assumption, $\psi_0$ vanishes on these elements,
hence on the ideal generated by them.
This implies $\psi_0$ is well-defined on $U(\mathcal{L}_0)$.
\end{proof}

\begin{lem}\label{kernel}
Let $\mathcal{J}_0$ be the left ideal of $U(\mathcal{L}_0)$ (the subspace
invariant under the multiplication from the left) generated by elements
of the form $\psi_0(a) - a$ for $a \in \mathcal{L}_0$.

Then $U(\mathcal{L}_0)/\mathcal{J}_0$ is nontrivial if and only if $\psi_0$ vanishes on
$[\mathcal{L}_0, \mathcal{L}_0]$. In this case $\mathcal{J}_0 = \ker \psi_0$ and
the quotient space is one-dimensional.
\end{lem}
\begin{proof}
If $\psi_0$ vanishes on $[\mathcal{L}_0, \mathcal{L}_0]$, then by lemma \ref{mult} $\psi_0$ extends
to $U(\mathcal{L}_0)$ and $\mathcal{J}_0$ is included in $\ker\psi_0$. Since $\psi_0$ is
nontrivial, $\ker\psi_0$ is nontrivial.

On the other hand, if $\psi_0$ doesn't vanish at $[\mathcal{L}_0, \mathcal{L}_0]$, then take
$x,y \in \mathcal{L}_0$ such that $\psi_0([x,y]) \neq 0$. Then it holds that
\begin{align*}
  [\left(\psi_0(x)-x\right), \left(\psi_0(y)-y\right)] = [x,y] \in \mathcal{J}_0, \\
  \psi_0([x,y]) - [x,y] \in \mathcal{J}_0.
\end{align*}
Hence $\mathcal{J}_0$ contains a nontrivial scalar and generates all.

To complete the proof, we only have to show that $\mathcal{J}_0 \supset \ker \psi_0$
since the other inclusion has been done. Therefore it is enough to show that
$\mathcal{J}_0$ has codimension 1 in $U(\mathcal{L}_0)$. This is a rephrasing
of the claim that any element of $U(\mathcal{L}_0)$
is equivalent to a scalar modulo $\mathcal{J}_0$.
This is easy to see since any element of $U(\mathcal{L}_0)$ is a linear combination
of tensor products $a_1\otimes a_2\otimes \cdots \otimes a_n$.
By definition there is an element
$a_1\otimes a_2\otimes \cdots \otimes(a_n-\psi_0(a_n))$ in $\mathcal{J}_0$,
therefore $a_1\otimes a_2\otimes \cdots \otimes a_n \equiv_{\mathcal{J}_0} 
a_1\otimes a_2\otimes \cdots \otimes \psi_0(a_n)$.
By repeating this procedure,
we see that every element of $U(\mathcal{L}_0)$ is equivalent to a scalar.
\end{proof}

In the following we assume that $\mathcal{L}$ is a *-Lie algebra with a decomposition
into Lie subalgebras
$\mathcal{L} = \mathcal{N}_- \oplus \mathcal{H} \oplus \mathcal{N}_+$, where
$(\mathcal{N}_+)^* = \mathcal{N}_-, (\mathcal{H})^* = \mathcal{H}$, and
$\mathcal{H}$ is commutative.

Let $\psi$ be a linear functional on $\mathcal{H} \oplus \mathcal{N}_+$ which vanishes on its commutator subalgebra.
In other words, $\psi$ is an element of $H^1(\mathcal{N_+} \oplus \mathcal{H}, \mathbb{C})$.
We will show that for any such $\psi$ we have a left module on $\mathcal{L}$.
Again let $U(\mathcal{L})$ be the universal enveloping algebra of $\mathcal{L}$.
It is naturally a left module on $\mathcal{L}$.
\begin{prp}\label{module}
Let $\mathcal{J}$ be the left ideal of $U(\mathcal{L})$ generated by
elements of the form $\psi(l_+) - l_+$, where $l_+ \in \mathcal{H} \oplus \mathcal{N}_+$.
The subspace $\mathcal{J}$ is a nontrivial submodule on $U(\mathcal{L})$.
\end{prp}
\begin{proof}
By the theorem of Poincar\`{e}-Birkhoff-Witt, it holds that
$U(\mathcal{L}) = U(\mathcal{N}_-) \otimes U(\mathcal{H}) \otimes U(\mathcal{N}_+)$.
By lemma \ref{kernel}, $\ker\psi$ has codimension one in
$U(\mathcal{H}) \otimes U(\mathcal{N}_+)$. It is easy to see that $\mathcal{J}$ takes the form
$U(\mathcal{N}_-) \otimes \ker\psi$, hence it is nontrivial.
\end{proof}

For a fixed $\psi$ we define the quotient module $V = U(\mathcal{L})/\mathcal{J}$.
Since $U(\mathcal{H}) \oplus U(\mathcal{N}_+) /\ker\psi$
is one dimensional, the module $V$ is linearly isomorphic to
$U(\mathcal{N}_-)$ and we identify them.
There is a specified vector $v$ which corresponds to $1 \in \C \subset U(\mathcal{N}_-)$
and, on $v$, an element $x$ of $\mathcal{H}\otimes\mathcal{N}_+$ acts as
$xv = \psi(x)v$.

\begin{exm}
The Virasoro algebra has the following decomposition:
\[
  \mathrm{Vir} = \mathcal{V}_- \oplus \mathcal{H} \oplus \mathcal{V}_+,
\]
where $\mathcal{V}_+ = \mathrm{span}\{L_n: n > 0\}$ and $\mathcal{H} = \mathrm{span}\{L_0, C\}$.
It is easy to see that the commutator subalgebra
$[\mathcal{H} \oplus \mathcal{V}_+, \mathcal{H} \oplus \mathcal{V}_+]$ is equal to
$\mathcal{V}_+$. According to proposition \ref{module}, we obtain a module
of $\mathrm{Vir}$ for any linear functional $\psi$ on $\mathcal{H} \oplus \mathcal{V}_+$
vanishing on $\mathcal{V}_+$. The linear functional $\psi$ is determined by the
two values $c := \psi(C)$ and $h := \psi(L_0)$.
It is well known that for some values of $c$ and $h$ we can define inner products on
these modules and these representations integrate to representations of the group
$\mathrm{Diff}(S^1)$ \cite{GW}.
\end{exm}

\begin{exm}
The *-Lie algebra $\mathcal{K}$ has the decomposition
\[
  \mathcal{K} = \mathcal{K}_+ \oplus \mathcal{H} \oplus \mathcal{K}_-,
\]
where $\mathcal{K}_+ = \mathrm{span}\{K_n: n > 0\}$ and
$\mathcal{H} = \mathrm{span}\{C\}$. It can be shown that
$H^1(\mathcal{K}_+ \oplus \mathcal{H}, \mathbb{C})$ is three dimensional
and an element $\psi$ in $H^1(\mathcal{K}_+ \oplus \mathcal{H}, \mathbb{C})$
takes the form
\[
  \psi(C) = c, \psi(K_n) = h+ih^\prime + \lambda n \text{ where } c, \lambda \in \mathbb{C}, h,h^\prime, \in \mathbb{R}.
\]
We denote this module on $\mathcal{K}$ by $V_{h+ih^\prime, c, \lambda}$.
If $c \in \mathbb{C}$,
$\psi(K_n) = h + ih^\prime\in \mathbb{C}$ and $\lambda = 0$
then the modules $V_{h+ih^\prime, c, 0}$ reduce to Verma modules on the Virasoro
algebra (see proposition \ref{extensiontovir}).
\end{exm}

Let us return to general cases.
 From now on we assume that $\psi$ is self-adjoint on $\mathcal{H}$
(namely, $\psi(h^*) = \overline{\psi(h)}$ for $h \in \mathcal{H}$).
Recall that $V$ is the quotient module $U(\mathcal{L})/\mathcal{J}$
as in the remark after proposition \ref{module}.
Our next task is to define a contravariant sesquilinear
form on $V$. Note that the *-operation extends naturally to $U(\mathcal{L})$.

We define a sesquilinear map on $V\times V$ ( = $U(\mathcal{N}_+)\times U(\mathcal{N}_+)$)
into $U(\mathcal{L})$ by
\[
  \alpha(L^-_1, L^-_2) = (L^-_2)^* \otimes L^-_1,  \text{ for } L^-_1, L^-_2 \in U(\mathcal{N}_-) = V.
\]

On the other hand, we can define a linear form $\beta$ on $U(\mathcal{L})$ using
the decomposition $U(\mathcal{N}_-) \otimes U(\mathcal{H}) \otimes U(\mathcal{N}_+)$, by
\[
  \beta(L_- \otimes H \otimes L_+) = \overline{\psi\left((L_-)^*\right)}\psi(H)\psi(L_+).
\]
It is easy to see that $\beta$ is self-adjoint since $\psi$ is self-adjoint on $\mathcal{H}$.

\begin{thm}
$\beta \circ \alpha := \gamma$ is contravariant.
\end{thm}
\begin{proof}
We have to show that for any $L \in \mathcal{L}$ it holds 
\[
  \gamma(L\otimes L^-_1, L^-_2) = \gamma(L^-_1, L^*\otimes L^-_2).
\]

As elements of $U(\mathcal{L})$, we have the following decompositions by
the Poincar\`{e}-Birkhoff-Witt theorem:
\begin{eqnarray}
  L\otimes L^-_1 &=& \sum_k L^-_k \otimes H_k \otimes L^+_k \nonumber \\
  (L^-_2)^* \otimes L^-_k &=& \sum_l L^-_{k,l} \otimes H_{k,l} \otimes L^+_{k,l} \label{expansion}\\
  H_k \otimes L^+_k \otimes H_{k,l} \otimes L^+_{k,l} &=& \sum_m
    H_{k,l,m} \otimes L^+_{k,l,m}, \nonumber
\end{eqnarray}
where elements in the decompositions are $L^-_k, L^-_{k,l} \in U(\mathcal{N}_-)$,
$H_k,H_{k,l},H_{k,l,m} \in U(\mathcal{H})$ and $L^+_k, L^+_{k,l}, L^+_{k,l,m} \in U(\mathcal{N}_+)$.

Now we calculate
\begin{eqnarray*}
  \gamma(L\otimes L^-_1,L^-_2) &=& \gamma\left(\sum_k L^-_k \psi(H_k \otimes L^+_k), L^-_2\right) \\
  &=& \sum_k \psi(H_k\otimes L^+_k) \beta\left((L^-_2)^*\otimes L^-_k\right) \\
\end{eqnarray*}
By substituting the expression in (\ref{expansion}) to $(L^-_2)^* \otimes L^-_k$, we have
\begin{eqnarray*}
\gamma(L\otimes L^-_1,L^-_2)
  &=& \sum_{k,l} \psi(H_k \otimes L^+_k) \overline{\psi\left((L^-_{k,l})^*\right)} \psi(H_{k,l} \otimes L^+_{k,l}) \\
  &=& \sum_{k,l} \overline{\psi\left((L^-_{k,l})^*\right)} \psi(H_{k,l} \otimes L^+_{k,l} \otimes H_k \otimes L^+_k) \\
\end{eqnarray*}
By substituting the expression in (\ref{expansion}) to $H_k \otimes L^+_k \otimes H_{k,l} \otimes L^+_{k,l}$,
\begin{eqnarray*}
\gamma(L\otimes L^-_1,L^-_2)
  &=& \sum_{k,l,m} \overline{\psi\left((L^-_{k,l})^*\right)} \psi(H_{k,l,m} \otimes L^+_{k,l,m}) \\
  &=& \beta\left(\sum_{k,l,m} L^-_{k,l} \otimes H_{k,l,m} \otimes L^+_{k,l,m}\right) \\
  &=& \beta\left(\sum_{k,l} L^-_{k,l} \otimes H_{k,l} \otimes L^+_{k,l} \otimes H_{k}\otimes L^+_{k}\right) \\
  &=& \beta\left(\sum_k (L^-_2)^* \otimes L^-_k \otimes H_{k}\otimes L^+_{k}\right) \\
  &=& \beta\left((L^-_2)^* \otimes L\otimes L^-_1\right).
\end{eqnarray*}
Similarly, in order to see $\beta\left((L^-_2)^* \otimes L\otimes L^-_1\right) = \gamma(L^-_1, (L)^*\otimes L^-_2)$
we need the following decompositions (we use same notations to save number of letters.).
\begin{eqnarray*}
  L^*\otimes L^-_2 &=& \sum_k L^-_k \otimes H_k \otimes L^+_k \\
  (L^-_k)^* \otimes L^-_1 &=& \sum_l L^-_{k,l} \otimes H_{k,l} \otimes L^+_{k,l} \\
  (L^-_{k,l})^* \otimes H_k \otimes L^+_k &=& \sum_m
    H_{k,l,m} \otimes L^+_{k,l,m},
\end{eqnarray*}
where elements in the decompositions are $L^-_k, L^-_{k,l} \in U(\mathcal{N}_-)$,
$H_k,H_{k,l},H_{k,l,m} \in U(\mathcal{H})$ and $L^+_k, L^+_{k,l}, L^+_{k,l,m} \in U(\mathcal{N}_+)$.
Now the final computation goes as follows.
\begin{eqnarray*}
  \gamma(L^-_1, L^*\otimes L^-_2) &=& \gamma\left(L^-_1, \sum_k L^-_k \otimes H_k \otimes L^+_k \right) \\
  &=& \sum_k \overline{\psi(H_k\otimes L^+_k)} \beta\left((L^-_k)^*\otimes L^-_1\right) \\
  &=& \sum_{k,l} \overline{\psi(H_k)\psi(L^+_k)}\overline{\psi\left((L^-_{k,l})^*\right)}\psi(H_{k,l})\psi(L^+_{k,l}) \\
  &=& \sum_{k,l} \overline{\psi\left((L^-_{k,l})^*\otimes H_k\otimes L^+_k\right)}\psi(H_{k,l})\psi(L^+_{k,l}) \\
  &=& \sum_{k,l,m} \overline{\psi(H_{k,l,m})\psi(L^+_{k,l,m})} \psi(H_{k,l})\psi(L^+_{k,l}). \\
\end{eqnarray*}
In the next step (and only here) we need the self-adjointness of $\psi$ on $\mathcal{H}$.
Continuing,
\begin{eqnarray*}
\gamma(L^-_1, L^*\otimes L^-_2) 
  &=& \sum_{k,l,m} \psi\left((H_{k,l,m})^*\right)\overline{\psi(L^+_{k,l,m})} \psi(H_{k,l})\psi(L^+_{k,l}) \\
  &=& \beta\left(\sum_{k,l,m} (L^+_{k,l,m})^*\otimes (H_{k,l,m})^*\otimes H_{k,l}\otimes L^+_{k,l}\right) \\
  &=& \beta\left(\sum_{k,l} (L^+_k)^*\otimes (H_k)^*\otimes L^-_{k,l}\otimes H_{k,l} \otimes L^+_{k,l} \right) \\
  &=& \beta\left(\sum_{k} (L^+_k)^*\otimes (H_k)^*\otimes (L^-_k)^*\otimes L^-_1 \right) \\
  &=& \beta\left((L^-_2)^*\otimes L\otimes L^-_1 \right).
\end{eqnarray*}
This completes the proof.
\end{proof}

In the case of $\mathrm{Vir}$, $c = \psi(C)$ and $h = \psi(L_0)$ must be real for the sesquilinear
form to be defined. For such $\psi$ it has been completely determined when the
sesquilinear forms are positive definite thanks to the Kac determinant formula \cite{KR}.

In the case of $\mathcal{K}$, the only condition for the existence of sesquilinear form
is that $\psi(C) \in \R$. Hence there are additional parameters $h^\prime \in \R, \lambda \in \C$
for generalized Verma modules $V_{h+ih^\prime, c, \lambda}$ on $\K$.

\subsection{Irreducibility of generalized Verma modules on $\K$}\label{irreducibility}
In this section, we completely determine for which values of $h+ih^\prime,c,\lambda$ the corresponding
generalized Verma modules on $\mathcal{K}$ are irreducible. The proof heavily
relies on the result of Feigin and Fuks \cite{FF} which has determined when the Verma
modules on the Virasoro algebra are irreducible. To utilize their result, we extend
the generalized Verma modules on $\mathcal{K}$ to (non-unitary) representations
of the Virasoro algebra.

Let $V_{h+ih^\prime,c,\lambda}$ be a generalized Verma module on $\mathcal{K}$
and $v$ be the corresponding lowest weight vector such that
\begin{align}\label{lowestweight}
K_nv = (h+ih^\prime+n\lambda)v \mbox{ for $n \ge 1$ and } Cv = cv.
\end{align}

First we observe that
\[
K_n \mapsto K_n - n\lambda I, C \mapsto C,
\]
where $I$ is the identity operator on $V_{h+ih^\prime,c,\lambda}$,
extends by linearity to a well-defined (non *-) representation
(on the same space $V_{h+ih^\prime+n\lambda}$)
of $\mathcal{K}$ (the proof is the same as that
of proposition \ref{auto}).
On the other hand, it is straightforward to see that this new representation
is equivalent to $V_{h+ih^\prime,c,0}$.
Irreducibility of a representation of an algebra is not changed even if we add
the identity operator to the set of operators.
Therefore the irreducibility of $V_{h+ih^\prime,c,\lambda}$ is equivalent to
that of $V_{h+ih^\prime,c,0}$
and we may restrict the consideration to the latter case. We denote it $V_{h+ih^\prime,c}$.

\begin{lem}
For any $w \in V_{h+ih^\prime,c}$ there is $N \in \mathbb{N}$ such that
$K_mw = K_nw$ for $m,n \ge N$.
\end{lem}
\begin{proof}
The module $V_{h+ih^\prime,c}$ is spanned by vectors $K_{n_1}\cdots K_{n_k}v$. We
will show the lemma by induction with respect to $k$.
If $w = v$, the lowest weight vector, then the lemma obviously holds with $N=1$, hence
the case $k=0$ is done.

Assume that the lemma holds for $w$ and put $\lim_m K_mw = w^\prime$
(here $\lim$ has nothing to do with any topology, but simply means that
``the equality holds for sufficiently large $m$'').
We will show that it also holds for $K_nw$.
Let us calculate
\begin{eqnarray*}
K_m K_n w &=& ([K_m,K_n]+K_n K_m)w \\
 &=& \left((m-n)K_{m+n} - mK_m + nK_n + K_nK_m\right)w,
\end{eqnarray*}
and for sufficiently large $m$ this is equal to
\[
(m-n)w^\prime - mw^\prime + nK_nw + K_nw^\prime = -nw^\prime + nK_nw+K_nw^\prime.
\]
and this does not depends on $m$.
\end{proof}

Let us define $Dw = \lim_m K_m w$. Then, it is clear that $D$ is a linear operator
on $V_{h+ih^\prime,c}$ and it holds $Dv = (h+ih^\prime)v$.

\begin{lem}
The following commutation relation holds:
\begin{align}
[D,K_n] = n(K_n - D).\label{fakehamiltonian}
\end{align}
\end{lem}
\begin{proof}
We only need to calculate
\begin{eqnarray*}
(DK_n-K_nD)w &=& \lim_m (K_mK_n-K_nK_m)w \\
 &=& \lim_m \left((m-n)K_{m+n} - mK_m + nK_n\right)w \\
 &=& n(K_n - D)w.
\end{eqnarray*}
\end{proof}

The relation (\ref{fakehamiltonian}) can be rewritten as $[K_n-D,-D] = n(K_n-D)$.

\begin{prp}\label{extensiontovir}
The representation of $\K$ on $V_{h+ih^\prime,c,0}$ extends to a representation of
$\vir$. This extension is the Verma module with $-h-ih^\prime,c$.
\end{prp}
\begin{proof}
We take a correspondence $L_0 \mapsto -D, L_n \mapsto K_n-D, C \mapsto C$.
Now that we know all the commutation relations between $D$ and $K_n$,
the confirmation that this correspondence is a representation is straightforward.

It is clear that the lowest weight vector is $v$ and $-Dv = (-h-ih^\prime)v$,
$(K_n-D)v = 0$ for $n \ge 0$, $Cv=cv$. We only have to show that
all the vectors of the form $(K_{n_1}-D)\cdots (K_{n_k}-D)v$,
where $n_1 \le \cdots \le n_k$, are linearly independent.
But this is clear from the fact that these vectors are eigenvectors
of $D$ and the fact that $\{K_{n_1}\cdots K_{n_k}v\}$ are independent by definition.
The former fact is shown by a straightforward induction.
\end{proof}

Here we remark that this extension of the representation does not change the
irreducibility. If the module on $\mathcal{K}$ is irreducible, then clearly
it is irreducible as a module on $\mathrm{Vir}$. On the other hand the operator $D$ above
is defined as the limit of $K_n$'s, hence if the module on $\mathcal{K}$ is
reducible then it is still reducible as a module on $\mathrm{Vir}$.

The following theorem is due to Feigin and Fuks \cite{FF}.
\begin{thm}
For $h,c \in \mathbb{C}$, the Verma module $V_{h,c}$ on the Virasoro
algebra is reducible if and only if there are natural numbers $\alpha, \beta$
such that
\begin{eqnarray*}
\Phi_{\alpha,\beta}(h,c)
 &:=& \left(h+\frac{1}{24}(\alpha^2-1)(c-13) + \frac{1}{2}(\alpha\beta-1)\right) \\
 & & \times \left(h+\frac{1}{24}(\beta^2-1)(c-13) + \frac{1}{2}(\alpha\beta-1)\right) \\
 & & + \frac{(\alpha^2-\beta^2)^2}{16} = 0.
\end{eqnarray*}
\end{thm}

The application of this to our case is now straightforward.
\begin{cor}
For $h,h^\prime \in \mathbb{R}$, $c, \lambda \in \mathbb{C}$, the generalized Verma
module $V_{h+ih^\prime,c,\lambda}$ on $\mathcal{K}$ is reducible if and only if
there are natural numbers $\alpha, \beta$ such that
\[
\Phi_{\alpha,\beta}(-h-ih^\prime,c) = 0.
\]
\end{cor}

\section{Endomorphisms of $\mathcal{K}$}\label{endomorphismsof}
This section is devoted to the study of *-endomorphisms of the algebra $\mathcal{K}$.
As in the case of automorphisms, endomorphisms of $\mathcal{K}$ are not natural
objects, but they are interesting from the viewpoint of representations.
We remarked before that any composition of a *-endomorphism and unitary representation
provides a unitary representation. In this way, we obtain a strange kind of representations
of $\mathcal{K}$. We will also have a rough classification of endomorphisms.

It is well known (for example, see \cite{LX}\cite{weiner}) that the following maps
are endomorphisms of the Virasoro algebra and they restrict to $\mathcal{K}$:
\begin{align*}
\delta_r (L_N) &= \frac{1}{r}L_{rn} + \frac{C}{24}\left(r-\frac{1}{r}\right),\\
\delta_r (C) &= rC,
\end{align*}
for any integer $r \in \mathbb{Z}$.

We have another type of *-endomorphisms of $\K$ parametrized by
a complex number $\alpha$. In the next section we will see that
these endomorphisms are related to
some unitary representation of $\mathrm{Diff(S^1)}_0$.

\begin{prp}\label{sigma}
Let $\alpha \in \mathbb{C}$ and $K$ be an element of $\mathcal{K}$ which satisfies
$[K,K^*] = -K-K^*$. Define
\begin{eqnarray*}
\sigma_\alpha(K_n) &=& \left(\frac{n^2+n}{2}\alpha + \frac{n^2-n}{2}\overline{\alpha} - \frac{n^2-n}{2}\right)K \\
                   & & + \left(\frac{n^2+n}{2}\alpha + \frac{n^2-n}{2}\overline{\alpha} - \frac{n^2+n}{2}\right)K^*, \\
\sigma_\alpha(C) &=& 0.
\end{eqnarray*}
Then $\sigma_\alpha$ extends to a *-endomorphism of $\mathcal{K}$ by linearity.
\begin{rmk}\label{sumkernel}
Examples of $K$ in this proposition are $K = K_1, -K_{-1}, -\frac{1}{6}K_2+\frac{2}{3}K_1$.
Since the image of $C$ is $0$, $\sigma_\alpha$ extends also to a
*-homomorphism of $\K_0$ into $\K$. Therefore, the kernel of $\sigma_\alpha$ is the direct
sum of $\ker \sigma_\alpha$ as a homomorphism of $\K_0$ and $\C C$.
\end{rmk}
\end{prp}
\begin{proof}
It is clear that $\sigma_\alpha$ preserves the *-operation. We only have to confirm that
it preserves commutation relations and this is done by straightforward calculations.
However, we will exhibit a clearer procedure.

Let us put $\beta = 3\alpha + \overline{\alpha} -1$. The definition of $\sigma_\alpha$ can be
rewritten as
\begin{align*}
\sigma_\alpha(K_n) = &\left(\frac{n^2-n}{2}\beta - (n^2-2n)\alpha\right)K \\
                    & + \left(\frac{n^2-n}{2}\beta - (n^2-2n)\alpha - n\right)K^*.
\end{align*}
If we put $\gamma_n = \frac{n^2-n}{2}\beta - (n^2-2n)\alpha$, this takes the form
$\sigma_\alpha(K_n) = \gamma_n K + (\gamma_n-n)K^*$.
Now it is easy to see that
\begin{align*}
[\sigma_\alpha(K_n), \sigma_\alpha(K_{-n})] &= [\gamma_n K + (\gamma_n-n)K^*, (\overline{\gamma_n}-n)K+\overline{\gamma_n}K^*] \\
 &= (-|\gamma_n|^2 + |\gamma_n - n|^2)(K+K^*) \\
 &= -n(2\mathrm{Re}\gamma_n-n)(K+K^*) \\
 &= -n\left(\sigma_\alpha(K_n) + \sigma_\alpha(K_{-n})\right).
\end{align*}
Next we calculate a general commutator, for $m \neq -n$,
\begin{eqnarray*}
& &[\sigma_\alpha(K_m), \sigma_\alpha(K_n)] \\
&=& \left(m\left(\frac{n^2-n}{2}\beta-(n^2-2n)\alpha-n\right)
                              -n\left(\frac{m^2-m}{2}\beta-(m^2-2m)\alpha-m\right)\right)\\
& &\times (K+K^*) \\
&=& \left(\frac{\beta}{2}-\alpha\right)(m^2n-mn^2)(K+K^*)
\end{eqnarray*}
On the other hand,
\begin{eqnarray*}
& &(m-n)\sigma_\alpha(K_{m+n}) - m\sigma_\alpha(K_m) + n\sigma_\alpha(K_n) \\
&=& \left((m-n)\gamma_{m+n} - m\gamma_m + n\gamma_n -(m-n)(m+n) + m^2 - n^2\right) \\
& & \times (K+K^*) \\
&=& \left(\frac{\beta}{2}-\alpha\right)(m^2n-mn^2)(K+K^*)
\end{eqnarray*}
and this completes the proof.
\end{proof}

\begin{prp}
Let us assume that $K+K^* \neq 0$.
If $\alpha \in \frac{1}{2} + i\mathbb{R}$, then $\ker(\sigma_\alpha)$ is
$\mathcal{K}_1\oplus \C C$ (see section \ref{commutators}). Otherwise, $\ker(\sigma_\alpha)$
is $\mathcal{K}_2\oplus \C C$.
\end{prp}
\begin{proof}
As we have noted in the remark \ref{sumkernel},
first we think $\sigma_\alpha$ as a homomorphism of $\K_0$.

By direct calculations, we have (see section \ref{basisfor}),
\begin{eqnarray*}
\rho(M^0_n) &=& \left(-(n+1)\alpha - n\overline{\alpha} + n\right)K
              + \left(-(n+1)\alpha - n\overline{\alpha} + n+1\right)K^*, \\
\rho(M^1_n) &=& (\alpha+\overline{\alpha}-1)(K+K^*), \\
\rho(M^2_n) &=& 0.
\end{eqnarray*}
The kernel of $\sigma_\alpha$ must be one of ideals in theorem \ref{idealstructure}.
 From this it is clear that $\ker(\sigma_\alpha)$ contains $\mathcal{K}_2$
and contains $\mathcal{K}_1$ if and only if $\mathrm{Re}\alpha = \frac{1}{2}$.

By the remark \ref{sumkernel}, the kernel of $\sigma_\alpha$ as a *-endomorphism
is $\K_1 \oplus \C C$ or $\K_2 \oplus \C C$, respectively.
\end{proof}

We have a partial classification of endomorphisms of $\K$.

\begin{prp}\label{endoclass}
If $\rho$ is a nontrivial *-endomorphism of $\mathcal{K}$, then the possibilities are:
\begin{enumerate}
\item $\rho = \sigma_\alpha$ with appropriate $K$ and $\alpha \in \frac{1}{2} + i\mathbb{R}$.
 In this case, $\ker(\rho) = \mathcal{K}_1 \oplus \C C$ and $\rho(K_1) = \alpha K + (\alpha-1)K^*$.
\item $\rho = \sigma_\alpha$ with appropriate $K$ and $\alpha \notin \frac{1}{2} + i\mathbb{R}$.
 In this case, $\ker(\rho) = \mathcal{K}_2 \oplus \C C$ and $\rho(K_1) = \alpha K + (\alpha-1)K^*$.
\item $\rho(K_1) = \sum_{i=1}^N a_iK_i + a_0C \in \mathcal{K}_+ \oplus \mathbb{C}C$,
 $\rho(K_2) = \sum_{i=1}^{2N} b_iK_i + b_0C
 \in \mathcal{K}_+ \oplus \mathbb{C}C$, where $\sum_{i = 1}^N a_i = \sum_{i = 1}^{2N} b_i = \frac{1}{N}$.
 In this case, $\ker(\rho) = \{0\}$.
\item $\rho(K_1) = \sum_{i=-N}^{-1} a_iK_i + a_0C \in \mathcal{K}_- \oplus \mathbb{C}C$, $\rho(K_2) =
 \sum_{i=2N}^{-1} b_iK_i + b_0C
 \in \mathcal{K}_+ \oplus \mathbb{C}C$, where $\sum_{i = N}^{-1} a_i = \sum_{i = 2N}^{-1} b_i = -\frac{1}{N}$.
 In this case, $\ker(\rho) = \{0\}$.
\item $\rho(K_n) = in\lambda C$ for some $\lambda \in \R$.
\end{enumerate}
\end{prp}
\begin{proof}
By lemma \ref{k1prop} and \ref{k2prop}, it takes the form $\rho(K_1) = \lambda K + \mu K^* + \nu C$,
$\rho(K_2) = \lambda^\prime K^\prime + \mu^\prime K^{\prime*} + \nu^\prime C$, where $K$ and $K^\prime$ are elements of
 $\mathcal{K}_+$.
Also by lemma \ref{proportional} with the commutation relation of $K_2$ and $K_{-1}$,
$K$ and $K^{\prime*}$ must be proportional.

If both of $\lambda$ and $\mu$ are nonzero, then also $K$ and $K^\prime$ must be proportional. By the commutation relation of $K_1$ and $K_{-1}$ we see that some scalar multiple of $K$ plus a central element (we call it temporarily $\tilde{K}$) satisfies $[\tilde{K},\tilde{K}^*] = -\tilde{K}-\tilde{K}^*$. Hence
from the beginning we may assume $[K,K^*] = -K-K^*+\kappa C$ for some $\kappa \in \mathbb{C}$.
Then again by the commutation relation, $\mu = \lambda -1$. Similarly, it holds $\mu^\prime = \lambda^\prime - 2$.
By the commutation relation of $K_2$ and $K_{-1}$ we see $\lambda^\prime = 3\lambda + \overline{\lambda} -1$.
Then this is exactly the case (1) or (2). It depends on the value of $\lambda$ whether it is (1) or (2).

Let one of $\lambda$ and $\mu$ be zero. By composing an automorphism $\tau$, we may assume $\mu = 0$
and we will show that we have the case (3). By the same argument of the beginning of theorem \ref{classauto},
$\rho(K_1)$ takes the form $\rho(K_1) = \sum_{i=1}^N a_iK_i + a_0C, \rho(K_2) = \sum_{j=1}^{2N} b_jK_j + b_0C$
and $\sum_{i=1}^N a_i = \frac{1}{N} = \sum_{i = 1}^{2N} b_i$.
Any finite set of $\rho(K_i)$'s is
linearly independent (by considering the highest or lowest terms of $\rho(K_i)$ in the
standard basis of $\mathcal{K}$) and we see $\ker(\rho) = \{0\}$.

If $\lambda = \mu = 0$, by the commutation relations \eqref{kcomm}, $\rho(K_2)$ must be mapped
to a central element. By the same argument as that of Lemma \ref{first}, $\rho$ is of the form
$\rho(K_n) = in\lambda C$.
\end{proof}

Let $\mathfrak{p}$ be the Lie algebra of the group generated by translations and
dilations in $\mathrm{Diff}(S^1)$. This algebra has a basis $\{T,D\}$ with the relation $[D,T] = T$
\cite{longo}\cite{lang}. Its complexification (which we denote again $\mathfrak{p}$)
is a *-Lie algebra with the *-operation $D^* = -D, T^* = -T$.
By setting $K = - D + iT$, we have $[K, K^*] = -K-K^*$.

\begin{lem}\label{algextension}
Any unitary representation $\varphi^\prime$ of $\mathfrak{p}$ produces a representation
$\varphi^\prime_1$ of $\mathcal{K}_0$ (or a representation of $\mathcal{K}$ with the central
charge $c = 0$).
\end{lem}
\begin{proof}
It suffices to set
\[
\varphi^\prime_1(K_n) = \frac{n^2+n}{2}\varphi'(K) + \frac{n^2-n}{2}\varphi'(K^*).
\]
We see that $\varphi^\prime_1$ preserves the commutation relations by the same computations
in the proof of proposition \ref{sigma} with $\alpha = 1$.
\end{proof}

\begin{rmk}
Any composition of a *-endomorphism and a unitary representation of $\K$
is again a unitary representation.
As we shall see in the next section, a composition of an endomorphism of type (1)
or (2) in proposition \ref{endoclass} and a lowest weight representation
gives rise to a strange representation (in the sense that they are ``localized at the point at
infinity''). On the other hand, a composition of the type (3) endomorphism
and a lowest weight representation
contains at least one lowest weight vector in the sense of subsection \ref{irreducibility},
equation (\ref{lowestweight}) which
is the lowest weight vector of the original representation, and the value of
$h+ih^\prime$ is changed to $\frac{1}{N}(h+ih^\prime)$.
If we start with the restriction to $\K$ of a unitary representation of $\vir$,
representations with ``complex energy'' (namely, $h^\prime \neq 0$) do not arise in this way.
\end{rmk}

\section{Some unitary representations of $B_0$}\label{someunitary}
In this section we will construct true (not projective) unitary representations
of $B_0$. Symmetries in physics are in general described by unitary projective
representations of a group \cite{schottenloher}. From this point of view,
one dimensional true representations are trivial, since they are equivalent
to the trivial representations as projective representations. Nevertheless,
we here exhibit a construction of a one dimensional representation. The author
believes that this reveals the big difference between $\mathrm{Diff}(S^1)$ and
$B_0$. In fact, $\mathrm{Diff}(S^1)$ does not admit any positive energy true representation
(see \cite{schottenloher}). This difference comes mainly from the fact that
$\mathrm{Diff(S^1)}$ is simple but $B_0$ is not simple.

We identify $B_0$ with a space of functions on $\mathbb{R}$ as in section \ref{thederived}.

\begin{prp}
For any $\lambda \in \R$ the map
\begin{eqnarray*}
\varphi: B_0 &\to&  S^1 \\
  f &\mapsto&  \exp(i\lambda \log f^\prime(0))
\end{eqnarray*}
is a (one-dimensional) unitary representation of $B_0$.
\end{prp}
\begin{proof}
Recall that $B_0$ is the group of orientation preserving,
$0$-stabilizing diffeomorphisms of $S^1$. By the identification with the function space,
the derivative of any element is everywhere (in particular at $\theta = 0$) positive,
hence the map is properly defined.

By the formula
\[
(f\circ g)^\prime(0) = f^\prime(0)\cdot g^\prime(0),
\]
we see the map $\varphi$ above is multiplicative.
\end{proof}

\begin{rmk}
This $\varphi$ is obviously irreducible and does not extend to $\mathrm{Diff}(S^1)$.
In fact, $\varphi$ is the integration of the one-dimensional representation of
corollary \ref{1dimrep}.
If $g \in B_0$ is localized on some closed interval which does
not include $0$, then $\varphi(g) = 1$. In this sense, $\varphi$ is ``localized at
the point at infinity''.
\end{rmk}

Next we need a general lemma.
\begin{lem}
Let $G$ be a group, $H$ a normal subgroup
of $G$ and $\pi$ the quotient map $G \to G/H$. 
Let F be a subgroup of $G$ such that $F \cap H = \{e\}$ and
$\pi(F) = G/H$. Then $G/H$ and $F$ are isomorphic by a canonical isomorphism $\gamma$
such that $\gamma \circ \pi|_F = \mathrm{id}$.
If $\varphi$ is a representation of $F$, it extends to a representation
$\tilde{\varphi} :=  \varphi \circ \gamma \circ \pi$ of $G$.
\end{lem}

Let $B_2 = \{g \in B_0: f^\prime(0) = 1, f^{\prime\prime}(0) = 0\}$.
It is easy to see that $B_2$ is a normal subgroup of $B_0$.

Let $G = B_0, H = B_2$ and $F = P$ be the subgroup generated by
dilations and translations. It is obvious that any element of $F$ can be written
as a product of a dilation and a translation. The derivative of a translation at point $0$ is
always $1$, whereas a nontrivial dilation has the derivative different from $1$ at $0$. From
this, the intersection of $F$ and $H$ must
be pure translations. But then, any element of this intersection must have
a vanishing second derivative at $0$. This implies that the intersection is trivial.

By a similar consideration, it is not difficult to see that $\pi(P) = \pi(B_0)$.
By the previous lemma, the unitary irreducible representation of
$F = P$ extends to a unitary irreducible representation of $B_0$ having
$B_2$ in the kernel.

Also this representation is ``localized at the point at infinity'', since if a
diffeomorphism is localized in a closed interval which does not contain $0$,
then it is an element of $B_2$ and hence mapped to the identity operator.

Summing up, we have the following.
\begin{thm}\label{canonicalextension}
Any unitary representation $\varphi$ of $P$ canonically extends to a representation
$\tilde{\varphi}$ of $B_0$ which is localized at the point at infinity.
\end{thm}

We describe the relation between this representation and the endomorphism
of $\mathcal{K}$ constructed in section \ref{endomorphismsof}.
The group $P$ admits a unique irreducible positive energy (which means that the generator
of translation is positive) true (not projective)
representation \cite{longo}.
This representation can be considered as the integration of several lowest weight
representations of the Lie algebra $\mathfrak{p}$ of $\mathcal{P}$.
In the following, we fix such a representation
of $\mathfrak{p}$ and extend it to $\mathcal{K}$. The representation space of $\mathfrak{p}$
is a dense subspace of the representation space of $P$ and it is the core of any generator
of one-parameter subgroup of $P$ (see \cite{longo}). Through $\tilde{\varphi}$, any
one-parameter subgroup $g_t$
of $B_0$ is first mapped to $P$ by $\gamma \circ \pi$ and then represented as a one-parameter
group of unitary operators. Hence any unbounded operator appearing here is in the
representation of $\mathfrak{p}$ explained above and there arise no problems of
domains or self-adjointness.

\begin{prp}
Let $\varphi$ be a unitary representation of the Lie group $P$, $\varphi^\prime$ be
the corresponding representation of the Lie algebra $\mathfrak{p}$
and $\varphi^\prime_1$ be the extension to $\mathcal{K}$ in proposition \ref{algextension},
then $\varphi^\prime_1$ integrates to $\tilde{\varphi}$ in the theorem \ref{canonicalextension}.
\end{prp}
\begin{proof}
The quotient group $B_0/B_2$ is isomorphic to $\mathbb{R}_+ \rtimes \mathbb{R}$
with the group operation:
\[
(X_1,X_2) \cdot (Y_1,Y_2) = (X_1Y_1,X_1Y_2+Y_1^2X_2), \mbox{ for } X_1,Y_1 \in \R_+, X_2,Y_2 \in \R.
\]
The isomorphism $\rho$ is given by $f \mapsto \left(f^\prime(0), f^{\prime\prime}(0)\right)$.

It's Lie algebra has the structure $\mathbb{R} \oplus \mathbb{R}$ with
\[
[(x_1,x_2),(y_1,y_2)] = (0,x_2y_1-x_1y_2)] \mbox{ for } x_1, x_2, y_1, y_2 \in \R.
\]
If $g^s$ is a one-parameter subgroup in $B_0$ with generator
$v$, then the corresponding element in the algebra is
$\rho^\prime(v) = \left(v^\prime(0), v^{\prime\prime}(0)\right)$, where
$\rho^\prime$ is the derivative of $\rho$.

The generator of the one-parameter subgroup of dilations $D_s(\theta)$ is
$\frac{1}{2}(K_1-K_1^*)(\theta) =: d_1(\theta) = \sin\theta$ and the generator of
translations $T_s(\theta)$
is $-\frac{i}{2}(K_1+K_1^*) =: t_1(\theta) = 1-\cos\theta$.
Thus $\rho^\prime(d_1) = (1,0)$ and $\rho^\prime(t_1) = (0,1)$.
Similarly, the generator
$\frac{1}{2}(K_n-K_n^*)(\theta) =: d_n(\theta) = \sin n\theta$ is mapped to
$(n,0)$ and $-\frac{i}{2}(K_n+K_n^*) =: t_n(\theta) = 1-\cos n\theta$ is mapped to $(0,n^2)$.
In short, it holds that $\rho^\prime(d_n) = n\rho^\prime(d_1),
\rho^\prime(t_n) = n^2\rho^\prime(t_1)$. Hence
these relations hold also for the derivative of $\tilde{\varphi}$,
namely $\tilde{\varphi}^\prime(d_n) = n\tilde{\varphi}^\prime(d_1),
\tilde{\varphi}^\prime(t_n) = n^2\tilde{\varphi}^\prime(t_1)$.

On the other hand, for $\varphi^\prime_1$ we have
\begin{align*}
\varphi^\prime_1\left(\frac{1}{2}(K_1-K_1^*)\right) &= \frac{1}{2}(K-K^*), \\
\varphi^\prime_1\left(-\frac{i}{2}(K_1+K_1^*)\right) &= -\frac{i}{2}(K+K^*), \\
\varphi^\prime_1\left(\frac{1}{2}(K_n-K_n^*)\right) &= \frac{n}{2}(K-K^*) =
 n\varphi^\prime_1\left(\frac{1}{2}(K_1-K_1^*)\right), \\
\varphi^\prime_1\left(-\frac{i}{2}(K_n+K_n^*)\right) &= -\frac{in^2}{2}(K+K^*)
 = n^2\varphi^\prime_1\left(-\frac{i}{2}(K_1+K_1^*)\right).
\end{align*}
 From this it is clear that $\varphi^\prime_1$ and $\tilde{\varphi}^\prime$ are equivalent,
since by definition $\varphi_1(d_1) = \tilde{\varphi}^\prime(d_1)$ and 
$\varphi_1(t_1) = \tilde{\varphi}^\prime(t_1)$
\end{proof}

As remarked before, there is a unique irreducible positive energy representation of $P$.
By the proposition above, it extends to an irreducible positive energy true
representation of $B_0$.

\section*{Acknowledgments}
I would like to thank Roberto Longo for his useful suggestions and constant support,
Paolo Camassa, John Elias Roberts and the referee of The International Journal of
Mathematics for their careful reading and correction of the manuscript.

A part of this work has been done during the author's visit to the Erwin Schr\"{o}dinger
Institute in Vienna for the program on Operator Algebras and Conformal Field Theory
in 2008. I am grateful to ESI for their support and hospitality.

\def\cprime{$'$}

\end{document}